\documentclass[superscriptaddress,pre,showpacs,preprintnumbers,amsmath,amssymb,aps]{revtex4}

\usepackage{graphicx}
\usepackage{dcolumn}
\usepackage{bm}
\usepackage{color}

\newcommand{\blue}[1]{\textcolor{blue}{#1}}

\begin{document}


\title{Factorization symmetry in lattice Boltzmann simulations}

\author{Ilya Karlin}\email{karlin@lav.mavt.ethz.ch}
\affiliation{Aerothermochemistry and Combustion Systems Lab, ETH Zurich, 8092 Zurich, Switzerland}
\affiliation {School of Engineering Sciences, University of Southampton, SO17 1BJ Southampton, UK}

\author{Shyam S. Chikatamarla}
\affiliation{Aerothermochemistry and Combustion Systems Lab, ETH Zurich, 8092 Zurich, Switzerland}

\author{Pietro Asinari}
\affiliation{Department of Energetics, Politecnico di Torino,
Corso Duca degli Abruzzi 24, Torino, Italy}

\date{\today}

\begin{abstract}
A non-perturbative algebraic theory of lattice Boltzmann method is developed based on a symmetry of a product.
It involves three steps: (i) Derivation of admissible lattices in one spatial dimension through a matching condition which imposes restricted extension of higher-order Gaussian moments,
(ii) Special quasi-equilibrium distribution function found analytically in closed form on the product-lattice in two and three spatial dimensions, and which proves factorization of quasi-equilibrium moments, and
(iii) Algebraic method of pruning based on a one-into-one relation between groups of discrete velocities and moments.
Two routes of constructing lattice Boltzmann equilibria are distinguished.
Present theory includes previously known limiting and special cases of lattices, and enables automated derivation of lattice Boltzmann models
from two-dimensional tables, by finding roots of one polynomial and solving a few linear systems.

\end{abstract}


\maketitle

\section{Introduction}
\label{sec:intro}

There were a few recent attempts \cite{CK_PRL2,CK09,Philippi06,Philippi08,Shan_JFM06,Shan_PhysicaD08,Nie08} to construct a theory of the lattice Boltzmann (LB) method - a modern approach to fluid dynamics \cite{Succi}.
This is due, in the first place, because LB models currently in use are not "sufficiently" Galilean invariant
(the feature that LB improved on from its predecessor, the lattice gas model, but failed to resolve completely).
Even though the Galilean non-invariance of current LB models was very well known right from the beginning \cite{Qian92,Qian93}, curing this drawback resisted for a long time.
Insufficient Galilean invariance of the very basic LB at a constant temperature is a precursor of many difficulties, in particular, in applications of LB to high Reynolds number hydrodynamics \cite{Hazi06,Prasianakis09}, multi-phase flows \cite{Wagner06} and compressible flows \cite{he98,Prasianakis07}.
It is quite well understood that the current "standard" LB models are too much constrained by the "small" number of the discrete velocities, and lattices with "more" velocities are required in order to overcome these limitations. However, early attempts to introduce lattices with more velocities were unsuccessful because of a severe numerical instabilities of the resulting LB schemes \cite{Alder95,Renda98,Qian98}.

Important progress was recently achieved in \cite{CK_PRL2,CK09}, where the construction of the higher-order LB was formulated as the construction of the entropy \cite{Karlin99}. In particular,  \cite{CK_PRL2,CK09} explained why some of the most obvious suggestions for higher-order lattices are bound to failure due to the fact that no entropy can be constructed for them. The entropy construction of Refs.\ \cite{CK_PRL2,CK09} has led to admissible lattices in three dimensions which enable LB models with better properties but derivation of such lattices (the procedure termed pruning in Ref.\ \cite{CK09}) remained a rather tedious search among large families of lattices. Apparently, some kind of simplicity was still missing at that stage, and a fully analytic approach to pruning is a challenging task. On the other hand, symmetry with respect to a group of rotations was invoked recently for a classification of isotropy of higher-order LB models (in two dimensions) \cite{Chen08}. However, the information about the isotropy of the higher-order lattices alone is insufficient if we want to address stability (or instability) and the form of the equilibrium on each specific lattice.

In this paper, we develop a theory of higher-order LB based on a symmetry of a product.
We remind that such a symmetry is deeply rooted in the classical kinetic theory since its beginning, the seminal Maxwell's derivation of the equilibrium of the three-dimensional. Isotropy (independence of the equilibrium on the direction) in Maxwell's famous derivation comes from the fact that the product of one-dimensional Maxwell distributions depends only on the isotropic quantity, the kinetic energy of the particles: $\exp(-v_x^2)\exp(-v_y^2)\exp(-v_z^2)=\exp(-\bm{v}\cdot\bm{v})$.
Our consideration of the lattice Boltzmann method is based on the products of one-dimensional functions.
The present theory of LB method is algebraic (rather than group-theoretic \cite{Chen08,Rubinstein08} or function-theoretic \cite{CK_PRL2,CK09}) and non-perturbative (it is not based on  polynomial expansions of the Maxwellian \cite{Philippi06,Philippi08,Shan_JFM06,Shan_PhysicaD08,Nie08}).
The latter is important for preserving the symmetry of the product, as we will see it below.
%
The resulting theory is remarkably constructive and simple, and consists of three major steps: The construction begins in one dimension where we identify admissible lattices (sec.\ \ref{sec:D1}).
At this step, we reveal the reference temperature (of the Maxwell distribution represented by the given one-dimensional lattice).
This information is then immediately transferred (sec.\ \ref{sec:UniQuE}) into three dimensions with the help of a special  unidirectional quasi-equilibrium on a "large" lattice formed by all possible direct products of one-dimensional velocities (Maxwell lattice) (for general issues related to quasi-equilibria see \cite{GKbook}).
The result of sec.\ \ref{sec:UniQuE} (see Eq.\ (\ref{eq:UniQuE}) below)  extends the product form onto the entire  quasi-equilibrium populations. The advantage of the  unidirectional quasi-equilibrium on product-lattices is twofold: It has a simple structure of the corresponding moment representation (see Eq.\ (\ref{eq:inverse}) below), and constructing the equilibrium is a mere substitution of the one-dimensional data for one-dimensional non-conserved moments. We distinguish between two routes to obtain the equilibrium for lattice Boltzmann models: The equilibration (minimization of the entropy function under constraints of local conservation) and the Maxwellization (promotion of Maxwell's equilibrium values for the non-conserved moments). The Maxwell lattice is an "ideal" lattice in three dimensions, it replicates all the information gained in one dimension. "Ideal" also means that the information about three dimensions is represented without correlations in the product-form (\ref{eq:UniQuE}).
Based on the results of sec.\ \ref{sec:UniQuE}, in sec.\ \ref{sec:Pruning} the analytical method of pruning is developed.
The main ingredient in this approach to pruning is the two-dimensional key-table which furnishes the one-into-one relation between groups of velocities and moments, and which is relatively easy to analyze even for large velocity sets.
The pruning algorithm is explained with the examples of the familiar D3Q27 lattice and the higher-order D3Q125 lattice.
In particular, the Maxwellization based on the pruning of the unidirectional quasi-equilibrium moment system, derives equilibrium distributions by solving linear algebraic systems.
Finally, the results are discussed in sec.\ \ref{sec:conclusion}.

\section{Maxwell lattices in one dimension}
\label{sec:D1}

Since our construction will be based on the one-dimensional lattices, it is important to sort it out right from the beginning which one-dimensional velocity sets are admissible, and which have to be rejected.
Therefore, we consider the one-dimensional sets of discrete velocities $V$, with  $Q$ the total number of the velocities (below, we consider $Q$ odd but same considerations apply also to $Q$ even).
The discrete velocities $v_{(i)}\in V$ are assumed integer-valued such that $v_{(i)}=i$.
The basic mirror symmetry of $V$ assumes that if $v_{(i)}\in V$ then also $v_{(-i)}\in V$, and thus stopped particles with $v_{(0)}=0$ are always included.
Corresponding populations are denoted $f_{(i)}$, and we use convenient normalization,  $f_{(i)}=\rho\varphi_{(i)}$.
Summation over discrete or integration over continuous velocities will be denoted as $\langle\dots\rangle$, thus $\rho=\langle f_{(i)}\rangle$.

Discrete velocities $V$ are so chosen as to reproduce the moments of the one-dimensional Maxwell distribution function,
\[f^{\rm M}_v=\rho \varphi^{\rm M}_v,\]
where
\begin{equation}
\label{eq:M}
\varphi^{\rm M}_v=\sqrt{\frac{\pi}{2T_0}}\exp\left\{-\frac{(v-u)^2}{2T_0}\right\}.
\end{equation}
Introducing
\[M^{\rm M}_{(n)}(T_0,u)=\langle \varphi^{\rm M}_v v^n\rangle,\]
these are
\begin{eqnarray*}
M^{\rm M}_{(0)}&=&1\ {\rm (normalization)},                                                      \\
M^{\rm M}_{(1)}&=&u\ {\rm (flow\ velocity)},                                                     \\
M^{\rm M}_{(2)}&=& T_0+ u^2=\Pi^{\rm M}\ {\rm (equilibrium\ pressure\ at\ unit\ density)},       \\
M^{\rm M}_{(3)}&=& 3T_0 u +  u^3=q^{\rm M}\ {\rm (equilibrium\ energy\ flux\ at\ unit\ density)},\\
M^{\rm M}_{(4)}&=& 3T_0^2+ 6T_0 u^2 + u^4=R^{\rm M},                                             \\
M^{\rm M}_{(5)}&=&15T_0^2 u + 10 T_0 u^3 + u^5,
\end{eqnarray*}
and so on, to which we refer as Maxwell's (M) moment relations.
\begin{table}
\begin{tabular}{|l|l|l|l|}
\hline
$Q$ & $V$ & Closure & $T_0$                                                     \\
\hline
$3$ &  $\{0,\pm 1\}$ & $v_{(i)}^3=v_{(i)}$& $1/3$                               \\
\hline
$5$& $\{0,\pm 1,\pm 3$\} &  $v_{(i)}^5=10v_{(i)}^3-9v_{(i)}$& $1\pm\sqrt{2/5}$  \\
\hline
$7$& $\{0,\pm 1,\pm 2, \pm 3\}$ & $v_{(i)}^7=14v_{(i)}^5-49v_{(i)}^3+36v_{(i)}$ & $0.697953$ \\
\hline
$9$& $\{0,\pm 1,\pm 2, \pm 3, \pm 5\}$ & $v_{(i)}^9=39v_{(i)}^7-399v_{(i)}^5+1261v_{(i)}^3-900v_{(i)}$ & $0.756081$, $2.175382$\\
\hline
$11$& $\{0,\pm 1,\pm 2, \pm 3, \pm 4 \pm 5\}$ & $v_{(i)}^{11}=55v_{(i)}^9-1023v_{(i)}^7+7645v_{(i)}^5-21076v_{(i)}^3+14400v_{(i)}$ & $ 1.062794$\\
\hline
\end{tabular}
\caption{One-dimensional Maxwell lattices with odd number of integer-valued velocities, $Q=3,5,7,9,11$.
Second column: Lattice vectors;
Third column: Closure relation, defining the reference temperature $T_0$ through the matching condition (fourth column).}
\label{table:minimal_lattices}
\end{table}

The first information revealed from the lattice is the reference temperature
 $T_0$ at which (a part of the) Maxwell's moment relations will be verified.
This is done with the help of the closure relation and the matching condition.
The closure relation for the set $V$ with $Q$ velocities ($Q$ odd) is a linear relation between the $Q$-th power of the velocities, $v_{(i)}^Q$, and the lower-order odd powers, starting with $v_{(i)}^{Q-2}$ and ending with $v_{(i)}$. Such a linear relation always exists, and reflects the fact that only $Q$ velocity polynomials, $1,v_{(i)},\dots,v_{(i)}^{Q-1}$ are linearly independent.
For example, for $V=\{0,\pm 1\}$ (D1Q3), the closure relation is $v_{i}^3=v_{(i)}$ (cube of any velocity from the D1Q3 set is the velocity itself),
for $V=\{0,\pm 1\,\pm 3\}$ (D1Q5) it is $v_{(i)}^5=10v_{(i)}^3-9v_{(i)}$, and so on.
The existence of the closure relation implies that the moment $M_{(Q)}=\langle \varphi_{(i)}v_{(i)}^Q\rangle$ cannot be assigned
at one's will, and that only the linear in $u$ term of this moment at equilibrium can be made consistent with the corresponding Maxwell's value $M_{(Q)}^{\rm M}$. This leads to the matching condition which decides about the reference temperature $T_0$.
For example, for D1Q3 the third-order moment $M_{(3)}=\langle \varphi_{(i)}v_{(i)}^3\rangle$ equals $M_{(3)}=u$ for any population set, equilibrium or not. On the other hand, the Maxwell's expression, $M_{(3)}^{\rm M}=3T_0 u +u^3$, contains also the cubic term $u^3$ which cannot be made consistent with the previous expression. Only the linear term can be matched, $3T_0u=u$, if the reference temperature is set to $T_0=1/3$.
Similarly, for D1Q5, $M_{(5)}^{\rm M}=15 \rho T_0^2u+O(u^3)$, thus, the matching condition for linear terms becomes $15T_0^2-30T_0+9=0$. The latter equation reveals two values of the reference temperature, $T_0=1\pm\sqrt{2/5}$.
This example also explains why the shortest D1Q5 lattice is $\{0,\pm 1, \pm 3\}$ and not
$\{0,\pm 1, \pm 2\}$: For the latter, the closure relation is $v_{(i)}^5=5v_{(i)}^3-4v_{(i)}$, and $T_0$ is found as a solution of $15T_0^2-15T_0+4=0$ which has no real-valued roots, and hence does not define any reference temperature. This procedure is immediately applicable to any lattice (Appendix \ref{appendix:generalQ}).
In Table \ref{table:minimal_lattices}, we collected Maxwell lattices with $Q=3,5,7,9,11$, together with the corresponding closure relations and the reference temperatures.

Once the reference temperature is revealed, we immediately derive the equilibrium values of the populations at $u=0$ and unit density (weights) $W_{(i)}$.

For this, we introduce the complete set of moments (at unit density):
\[M_{(0)}=1=\langle \varphi_{(i)}\rangle,\ M_{(1)}=\langle \varphi_{(i)} v_{(i)}\rangle,\ \dots,\ M_{(Q-1)}=\langle \varphi_{(i)}v_{(i)}^{(Q-1)}\rangle.
\]
Denote ${\cal M}=\{M_{(1)},\dots,M_{(Q-1)}\}$ the totality of the moments, excluding $M_{(0)}=1$, and $\Omega$ the set of their values at which the solution to the latter $Q\times Q$ linear system is positive. This solution $\varphi_{(i)}({\cal M})$ is always easily found from the above  $Q\times Q$ linear system, and we denote
\[f^*_{(i)}=\rho\varphi_{(i)}({\cal M}).\]
For example, for D1Q3, functions $\varphi_{(i)}(u,\Pi)$ are found as the solution to a $3\times 3$ linear system,
$\varphi_{(0)}= (1-\Pi)$,
$\varphi_{(\pm 1)}=(1/2)(\Pi\pm u)$,
where $u=M_{(1)}$ and $\Pi=M_{(2)}$, while $\Omega=\{u,\Pi: 0\le \Pi\le 1, |u|\le\Pi\}$. For the $D1Q5$, we need two more moments of order three and four, $q=\langle \varphi_{(i)}v_{(i)}^3\rangle$ and $R=\langle \varphi_{(i)}v_{(i)}^4\rangle$, and thus
\begin{eqnarray*}
\varphi_{(0)} &=& 1-\Pi+\frac{1}{9}(R-\Pi),\\
\varphi_{(\pm1)} &=&\frac{1}{16}\left[\pm(9u-q)-R+9\Pi\right],\\
\varphi_{(\pm3)}&=&\frac{1}{48}\left[\pm(q-u)+\frac{1}{3}(R-\Pi)\right],
\end{eqnarray*}
and so forth.
In order to reveal the weights, we substitute the equilibrium values of the corresponding moments at $u=0$ into the above formulas for $\varphi_{(i)}$, at $T_0$ already available, to derive
\[W_{(i)}=\varphi_{(i)}({\cal M}^{\rm M}(T_0,0)).\]
This gives $W_{(0)}=2/3$, $W_{(\pm 1)}=1/6$ for D1Q3, $W_{(0)}=(4/45)\left(4 + \sqrt{10}\right)$, $W_{(\pm 1)}=(3/80) \left(8 - \sqrt{10}\right)$, $W_{(\pm 3)}=(1/720)\left(16 - 5 \sqrt{10}\right)$ for D1Q5 (at $T_0=1-\sqrt{2/5}$) and so on.

Once the weights and the reference temperature are derived, we can immediately proceed with the evaluation of the equilibrium populations. There are two options:

\begin{enumerate}
\item[(i)]{\it Equilibration.} The weights $W_{(i)}>0$ define the entropy function
$H=\langle f_{(i)}\ln(f_{(i)}/W_{(i)})\rangle$.
 The equilibrium populations
$f_{(i)}^{\rm E}=\rho\varphi_{(i)}^{\rm E}$ are defined as the minimum of $H$, conditioned by density $\rho$ and velocity $u$.
Let us distinguish between the velocity $u$ and the higher-order moments by writing
\[{\cal M}=\{u,{\cal N}\},\]
so that
\[{\cal N}=\{M_{(2)},\dots, M_{(Q-1)}\}.\]
The above functions
$\varphi_{(i)}(u,{\cal N})$ are substituted into $H$ to give $H(\rho,u,{\cal N})=\rho\ln \rho+\rho\tilde{H}(u,{\cal N})$, where
\[\tilde{H}(u,{\cal N})=\left\langle\varphi_{(i)}(u,{\cal N})\ln\left(\varphi_{(i)}(u,{\cal N})/W_{(i)}\right)\right\rangle.\]
%
The equilibrium is found from the equations,
\[\frac{\partial \tilde{H}(u,{\cal N})}{\partial M_{(2)}}=0,\ \dots,\ \frac{\partial \tilde{H}(u,{\cal N})}{\partial M_{(Q-1)}}=0.\]
These equations define the equilibrium solution ${\cal N}^{\rm E}={\cal N}^{\rm E}(u)$. Exact solution is available (so far) only for the D1Q3,   where ${\cal N}$ consists of the pressure $\Pi$ only; then $\Pi^{\rm E}=(1/3)(2\sqrt{1+3u_{\alpha}^2}-1)$. In other cases, various solution procedures  can be readily applied to get approximations to ${\cal N}^{\rm E}$ \cite{CK09}. The equilibrium is thus
\[f_{(i)}^{\rm E}=\rho\varphi_{(i)}(u,{\cal N}^{\rm E}),
\]
where ${\cal N}^{\rm E}$ is exact or approximate solution to the extremum condition.

\item[(ii)]{\it Maxwellization.} Alternatively, we can promote Maxwell's expressions of the moments ${\cal N}^{\rm M}(T_0,u)$ to derive a different set of equilibrium populations,
    \[f_{(i)}^{\rm M}=\rho\varphi_{(i)}(u,{\cal N}^{\rm M}).\]
For example, the Maxwellization of the D1Q5 model is accomplished upon substitution of $\Pi^{\rm M}$, $q^{\rm M}$ and $R^{\rm M}$ in the above expressions for $\varphi_{(0)}$, $\varphi_{(\pm 1)}$ and $\varphi_{(\pm 3)}$:

\begin{align}
\begin{split}
f^{\rm M}_{(0)} &= \rho\left\{1-(T_0+u^2)+\frac{1}{9}\left[(3T_0^2+ 6T_0 u^2 + u^4)-(T_0+ u^2)\right]\right\},\nonumber\\
f^{\rm M}_{(\pm1)}&=\frac{1}{16}\rho\left\{\pm[9u-(3T_0u+u^3)]-(3T_0^2+ 6T_0 u^2 + u^4)+9(T_0+ u^2)\right\},\nonumber\\
f^{\rm M}_{(\pm3)}&=\frac{1}{48}\rho\left\{\pm[(3T_0u+u^3)-u]+\frac{1}{3}[(3T_0^2+ 6T_0 u^2 + u^4)-(T_0+ u^2)]\right\},\nonumber
\end{split}
\end{align}
with $T_0=1-\sqrt{2/5}$.

\end{enumerate}

\begin{figure}[ht]
  \begin{center}
    \includegraphics[width=0.5\textwidth]{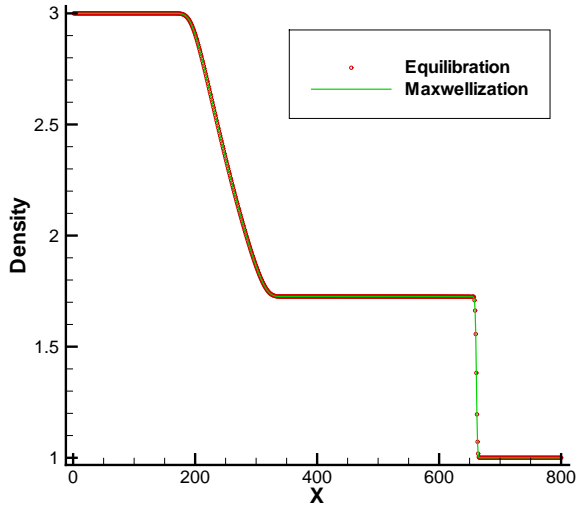}
  \end{center}
  \caption{(Color online) Shock tube test: Comparison of the Maxwellization on the D1Q5 lattice $V=\{0,\pm 1,\pm 3\}$) (line) with the equilibration of Ref.\ \cite{CK_PRL2} on the same lattice (symbol).
  Initial condition for the simulation was a density step,
$\rho=3.0$ for $x<L/2$ ($L$ being the length of domain),
$\rho=1.0$ for $x>L/2$ (same as in \cite{Qian98,CK_PRL2}). The snapshot of the density profile corresponds to kinematic viscosity
$\nu=0.138$ (the LBGK model of Ref.\ \cite{Qian98} on the inadmissible lattice $V=\{0,\pm 1,\pm 2\}$ is unstable at this viscosity and is not shown).}
  \label{Fig1}
\end{figure}

Maxwellization is easier than equilibration since functions ${\cal N}^{\rm M}$ are known from the one-dimensional Maxwellian $f_v^{\rm M}$.
In a contrast to the continuous velocity case where $f_v^{\rm E}=f_{v}^{\rm M}$,  Maxwellization is not the same as the equilibration on the lattice.
In order to illustrate this point, we present a comparison of the lattice Bhatnagar-Gross-Krook (LBGK) simulation of a one-dimensional shock propagation with two different equilibria.
In Ref.\ \cite{CK_PRL2}, it was shown that the LBGK model on the Maxwell D1Q5 lattice $V=\{0,\pm 1,\pm 3\}$, with the equilibrium constructed by the equilibration procedure (that is, via the entropy minimization) is superior in terms of numerical stability to the LBGK model on the inadmissible lattice  $V=\{0,\pm1,\pm 2\}$ of Ref.\ \cite{Qian98}.
In Fig.\ \ref{Fig1}, we present the result of the same simulation for the LBGK model with the equilibrium obtained by Maxwellization. Both simulations, with $f^{\rm E}$ \cite{CK_PRL2} and $f^{\rm M}$ (present) agree well with each other, and show the same stability properties.

Thus, the one-dimensional decoding is complete, we have derived reference temperatures and weights for an arbitrary one-dimensional velocity set just from the lattice itself.
In the next step  we are going to transmit the one-dimensional information into three dimensions. We close this section with a few comments:
\begin{itemize}
\item The reference temperatures for the Maxwell lattices collected in Table \ref{table:minimal_lattices}, and the corresponding weights, coincide with those found in Refs. \cite{CK_PRL2,CK09} using the entropy construction \cite{Karlin99}.
    The entropy construction derives the weights and the reference temperatures by comparing higher powers of velocity of ${\cal M}^{\rm E}$ to Maxwell moments.
    However, the present derivation via closure relation and matching condition is more direct and simpler. While the coincidence of the results obtained by two methods is quite remarkable, and suggests that the two approaches may be equivalent, the full proof of this statement is not available at the time of this writing, and is left for a further study.

\item It should be stressed that the three-velocity case (the basis of the "standard" LB models) is an exception: any set $V=\{0,\pm r\}$ is Maxwellian (the corresponding closure relation, $v_{(i)}^3=r^2v_{(i)}$, results - through the matching condition - in only a trivial re-scaling of the reference temperature, $T_0=r^2/3$). With $Q>3$, by far not every lattice is Maxwellian (see the above example of $V=\{0,\pm 1,\pm 2\}$ and Appendix \ref{appendix:D1Q5}). The above concept of  Maxwellization applies exclusively to the Maxwell lattices.
    Additional comments on Maxwell lattices and matching condition will be given in sec.\ \ref{sec:conclusion}.

\item
The two values for the reference temperature for $Q=5,9$ (Table \ref{table:minimal_lattices}) correspond to a Gaussian-like shape of the weights ($W_{|i|}\le W_{|j|}$ if $|i|>|j|$) for smaller $T_0$, and to a non-Gaussian shape for larger $T_0$  (cf.\ Ref.\ \cite{CK09}). Below, we consider $T_0$ corresponding to the Gaussian-like case in all the examples.
\end{itemize}

\section{Maxwell lattices in three dimensions and Unidirectional Quasi-Equilibrium}
\label{sec:UniQuE}

\subsection*{Unidirectional Quasi-Equilibrium}

In three dimensions, we first construct the product-lattice (or Maxwell lattice), induced by the one-dimensional Maxwell velocity set $V$, that is,
\begin{enumerate}
\item[(i)] The velocities are direct products of one-dimensional velocities,
\[v_{(i,j,k)}=(v_{(i)},v_{(j)},v_{(k)}),\]
\item[(ii)] Corresponding weights are algebraic products of the one-dimensional weights,
\[W_{(i,j,k)}=W_{(i)}W_{(j)}W_{(k)}.\]
\end{enumerate}
The  entropy  on the product-lattices is defined as
\[H=\left\langle f_{(i,j,k)}\ln\left(\frac{f_{(i,j,k)}}{W_{(i)}W_{(j)}W_{(k)}}\right)\right\rangle.\]
%
Moreover, the density is defined in the usual way,
$\rho=\langle f_{(i,j,k)}\rangle$, and we introduce a set of special unidirectional moments ${\cal M}_{\alpha}$ defined as
\begin{equation}\label{eq:UDmoments}
\rho M_{x(n)}=\langle f_{(i,j,k)}v_{(i)}^n\rangle,\ \rho M_{y(n)}=\langle f_{(i,j,k)}v_{(j)}^n\rangle,\ \rho M_{z(n)}=\langle f_{(i,j,k)}v_{(k)}^n\rangle,\ n=1,\dots,Q-1.
\end{equation}
Note that $M_{\alpha(1)}=u_{\alpha}$ are the components of the three-dimensional velocity, while the rest of the unidirectional moments are the
diagonal components of the corresponding tensors. For instance, $M_{\alpha(2)}$ are the diagonal components of the pressure tensor (at unit density), $M_{\alpha(3)}$ - of the third-order moment tensor etc.

Using  (\ref{eq:UDmoments}), we define a special unidirectional quasi-equilibrium state (UniQuE) as
the minimizer of the entropy function under the constraints imposed by fixed density and fixed unidirectional moments (\ref{eq:UDmoments}). That is,
UniQuE populations $f^*(\rho,{\cal M}_{x},{\cal M}_{y},{\cal M}_{z})$ are defined as the solution to the variational problem,
\begin{equation}
\label{eq:Hmin}
H\to \min,\  \langle f_{(i,j,k)}\rangle=\rho,\ \langle f_{(i,j,k)}v_{(i)}^n\rangle=\rho M_{x(n)},\ \langle f_{(i,j,k)}v_{(j)}^n\rangle=\rho M_{y(n)},\ \langle f_{(i,j,k)}v_{(k)}^n\rangle=\rho M_{z(n)},\ n=1,\dots,Q-1.
\end{equation}
The central result of this section is given by the following Theorem:

Solution to the conditional minimization problem (\ref{eq:Hmin}) is explicitly given by the formula
 \begin{equation}
\label{eq:UniQuE}
f_{(i,j,k)}^*=\rho\varphi_{(i)}({\cal M}_x)\varphi_{(j)}({\cal M}_y)\varphi_{(k)}({\cal M}_z),
\end{equation}
where the positive one-dimensional populations $\varphi_{(q)}({\cal M}_{\alpha})$ are defined by solving the one-dimensional moment system.

%
%

%

To prove this (see Appendix \ref{appendix:theoremUniQuE}), it is sufficient to notice that the solution to the minimization problem in terms of the Lagrange multipliers reduces to three decoupled one-dimensional problems of the form, $\langle \varphi_{(i)}\rangle=1$, $\langle \varphi_{(i)}v_{(i)}^{n}\rangle=M_{x(n)}$, and similarly for $y,z$. Solution of each of these problems is given by the unidirectional functions $\varphi({\cal M}_{\alpha})$ discussed in sec.\ \ref{sec:D1}.

UniQuE (\ref{eq:UniQuE}) is a family of populations defined by $3Q-2$ parameters in the $Q^3$-dimensional space, and is a fully factorized population: In order to construct (\ref{eq:UniQuE}), we plug ${\cal M}_{\alpha}$ instead of ${\cal M}$ in the one-dimensional functions $\varphi_{(i)}({\cal M})$, and multiply results for various $i$ and $\alpha$.

The above theorem about UniQuE applies to any Maxwell lattice.
We note in passing that special versions of UniQuE for the two-dimensional D2Q9 lattice was constructed in \cite{Prasianakis07} and \cite{Asinari09}, and for the D3Q27 - in \cite{Nikos_thesis} from the direct minimization of entropy. UniQuE (\ref{eq:UniQuE}) is the most crucial element in passing the information to three dimensions.
Note that, in general, it is impossible to find closed-form expressions for a quasi-equilibrium which minimizes the entropy under arbitrary constraints. UniQuE is the exceptional case because the solution is induced by the one-dimensional solutions which are explicitly known. This is possible only with the special choice of the constraints (unidirectional moments),  and  only on Maxwell lattices. This has a few immediate implications, two of which will be mentioned now.

\subsection*{Moment representation}

The product-lattice generated by $Q$ one-dimensional velocity vectors is characterized by
$Q^3$ linearly independent moments,
\[\rho M_{lmn}=\langle f_{(i,j,k)}v_{(i)}^lv_{(j)}^m v_{(k)}^n\rangle,\ l,m,n\in\{0,\dots, Q-1\}.\]
On the other hand, UniQuE is fully described by only $3Q-2$ moments (density and unidirectional moments $M_{l00}=M_{x(l)}$ etc).
Thus, the rest of the moments
become functions of density and unidirectional moments when evaluated on the UniQuE (\ref{eq:UniQuE}).
Evaluation is straightforward thanks to the product-form of the latter:
\begin{equation}\label{eq:inverse}
M_{lmn}^*= M_{x(l)}M_{y(m)}M_{z(m)}.
\end{equation}
Thus, the moment representation of UniQuE  (\ref{eq:inverse})
is a simple algebraic rule: One considers all possible products of functions $M_{\alpha(p)}$ with different spatial index $\alpha$, times the density $\rho$, where the number of functions in each such product does not exceed three. Example of the UniQuE moment system (\ref{eq:inverse}) for D3Q27 Maxwell lattice is given below in Table \ref{table:PP}.
Finally, since the moment and the population representations are equivalent to each other, we can now read (\ref{eq:inverse}) "from the right to the left" and say that it defines UniQuE upon inverting the $Q^3\times Q^3$ linear system (\ref{eq:inverse}) with the specified right hand side. This remark will be important later when we will consider sub-lattices of the product-lattice.

\subsection*{Equilibration}
The term "quasi-equilibrium" in the notion of UniQuE means that it is "less equilibrated" than the equilibrium.
The equilibrium (at the fixed reference temperature) minimizes entropy under fixed density and velocity $u_{\alpha}=M_{\alpha(1)}$.
Let us distinguish between the velocity $u_{\alpha}$ and the higher-order moments by writing  ${\cal M}_{\alpha}=\{u_{\alpha},{\cal N}_{\alpha}\}$. The above theorem about UniQuE  implies that the following two routes to equilibrium are equivalent:
\begin{itemize}
\item  The direct equilibration through minimization of $H$ under fixed $\rho$ and $u_{\alpha}$, and
\item The two-step equilibration, of which the first step is the "quasi-equilibration" by minimizing $H$ under fixed $\rho$ and ${\cal M}_{\alpha}$ (resulting in UniQuE), followed by the second equilibration step during which the UniQuE entropy
$H^*=\rho\ln\rho +\rho[\tilde{H}(u_x,{\cal N}_x)+\tilde{H}(u_y,{\cal N}_y)+\tilde{H}(u_z,{\cal N}_z)]$
 is minimized with respect to ${\cal N}_{\alpha}$ under fixed $\rho$ and $u_{\alpha}$, $\alpha=x,y,z$.
\end{itemize}
It is obvious from the product-form of UniQuE (\ref{eq:UniQuE}) that the second minimization reduces to the one-dimensional equilibration of sec.\ \ref{sec:D1}, and thus
\begin{equation}
\label{eq:EEQ}
f_{(i,j.k)}^{\rm E}=\rho\varphi_{(i)}(u_x,{\cal N}_x^{\rm E})\varphi_{(j)}(u_y,{\cal N}_y^{\rm E})\varphi_{(k)}(u_z,{\cal N}_z^{\rm E}).
\end{equation}
This again requires only the input  from the one-dimensional lattice (functions ${\cal N}_{\alpha}^{\rm E}$).
In other words, the UniQuE becomes equilibrium when the equilibrium values of the unidirectional moments are substituted into (\ref{eq:UniQuE}).
A few comments are in order:
The lattice Boltzmann equilibria on the product-lattices are constructed in such a way that the higher-order tensorial moments of a certain order render isotropic (to a certain order in the powers of the velocity components $u_{\alpha}$) \cite{CK09}. On the contrary, the special quasi-equilibria considered above are  anisotropic (their construction is based explicitly on a fixed Cartesian system of coordinates which is manifest in our choice of the parameters, the unidirectional moments).
Yet, the evaluation of these anisotropic moments (which are typically the diagonal components of the corresponding higher-order tensors) at the equilibrium renders the same degree of isotropy for the entire moment tensors at the equilibrium. Or, in other words, the  control (bringing to the equilibrium) over just the diagonal components of moment tensors is sufficient to control the entire tensors (including various off-diagonal components which are not explicitly targeted in the construction of the quasi-equilibrium).
This fully corresponds to Maxwell's argument on how the equilibrium in the three-dimensional gas become isotropic based on the independence of the three directions.

\subsection*{Maxwellization}

Same as in sec.\ \ref{sec:D1}, there is a different route to define the equilibrium on the product lattice by simply plugging in Maxwell's values ${\cal N}^{\rm M}(T_0,u)$ into UniQuE (\ref{eq:UniQuE}) to get a three-dimensional Maxwellization,
\begin{equation}\label{eq:PEQ}
f^{\rm M}_{(i,j.k)}=\rho\varphi_{(i)}(u_x,{\cal N}_x^{\rm M})\varphi_{(j)}(u_y,{\cal N}_y^{\rm M})\varphi_{(k)}(u_z,{\cal N}_z^{\rm M}).
\end{equation}
Note that (\ref{eq:PEQ}) is not the same as (\ref{eq:EEQ}). Moreover, (\ref{eq:PEQ}) differs also from the standard polynomial equilibrium on the product-lattices (for example, for the D3Q27, (\ref{eq:PEQ}) is a polynomial of the order six, while it is a second-order polynomial in the standard LB model). As an illustration, we collected all the populations mentioned so far (UniQuE, equilibration and Maxwellization) for the D3Q27 in Appendix \ref{appendix:D3Q27}.

\subsection*{Discussion}

Thus, the transmission of the one-dimensional information to three dimensions is now completed for the Maxwell lattice.
Arguably, this is a transmission "without errors", all the information about the Maxwell's relations collected for the one-dimensional distribution is manifest in the three dimensions once the product-lattice is used.
For example, the Maxwellization on the Maxwell lattices  (\ref{eq:PEQ}) recovers $Q^3$ moments of the three-dimensional Maxwellian:
\begin{equation}\label{eq:MaxMom}
M_{lmn}^{\rm M}= M_{x(l)}^{\rm M}M_{y(m)}^{\rm M}M_{z(m)}^{\rm M}.
\end{equation}
Moment relations (\ref{eq:MaxMom}) set the maximal possible accuracy achievable on the Maxwell lattice (for example, the moment system as recovered by the kinetic equation
$\partial_t f+v\cdot \nabla f=-(1/\tau)(f-f^{\rm M})$ is the closest approximate to a truncated moment equations system of the Boltzmann equation with the Bhatnagar-Gross-Krook collision operator).
Note that, in general, product-form of equilibria such as (\ref{eq:PEQ}) or (\ref{eq:EEQ}) should be preferred in LB computations \cite{CK_PRL1}. Whereas LB equilibria found by other methods (in particular, those using a polynomial expansion of the Maxwellian and quadrature approximations \cite{Philippi06,Philippi08,Shan_JFM06,Nie08}) can be recovered upon a further expansion and neglect of higher-order terms in (\ref{eq:PEQ}) or (\ref{eq:EEQ}), this discussion remains out of scope of the present paper since the method used here unambiguously results in the product-forms (\ref{eq:PEQ}) and (\ref{eq:EEQ}).

The drawback, however, is that the number of the velocities needed for this "error-free" transmission grows as $Q^3$ which becomes a large number.
Therefore, we need to consider an "incomplete" transmission by sacrificing some of the moments and reducing the number of velocities accordingly (pruning).
Above, we have remarked that UniQuE of the product-lattice  can be computed from the full $Q^3\times Q^3$ linear moment relations (\ref{eq:inverse}).
However, if we consider a part of the moment system (\ref{eq:inverse}) including moments of primary importance to the hydrodynamics only, this computation can be accomplished with a lesser number of the populations, or, equivalently, with a lesser number of the lattice velocities. In view of a large number of different moments, how to do this in a systematic fashion?
The answer to this question is central to the present theory, and will be given in the next section.

\section{Pruning and sub-Maxwell lattices}
\label{sec:Pruning}

\subsection{Backbone moments and sub-Maxwell lattices from Key-Table}
\label{sec:BBM}

\begin{table}
\begin{tabular}{|l|l|l|l|l|l|l|}
\hline
$0$& $2$ & $4$ & $6$ & $8$ & $10$ & $12$ \\
\hline
$1$ & $\Pi_{\alpha}$ & $\Pi_{\alpha}\Pi_{\beta}$ ($\alpha\ne\beta$)                             & $\Pi_x\Pi_y\Pi_z$                       &  &  & \\
&   &                &      $R_{\alpha}$              & $\Pi_{\alpha}\Pi_{\beta}R_{\gamma}$ ($\alpha\ne\beta\ne\gamma$) & $\Pi_{\alpha}R_{\beta}R_{\gamma}$ ($\alpha\ne\beta\ne\gamma$) & $R_xR_yR_z$ \\
&              &  & $\Pi_{\alpha}R_{\beta}$ ($\alpha\ne\beta$)& $R_{\alpha}R_{\beta}$ ($\alpha\ne\beta$)& & \\
\hline
\end{tabular}
\caption{Backbone moments of the D3Q125 UniQuE system (\ref{eq:inverse}) arranged in columns according to their order. Upper left corner displays the backbone moments of the D3Q27 (see Eq.\ (\ref{eq:D3Q27})).}
\label{table:BBMD3Q125}
\end{table}

Looking back at (\ref{eq:inverse}), we notice that only even-order moments give a non-vanishing contribution to this system at the equilibrium at velocity equal to zero. Indeed, the odd-order moments such as $u_{\alpha}$, $Q^{\rm E}_{\alpha\beta\gamma}$ or  $Q^{\rm M}_{\alpha\beta\gamma}$, etc. all vanish at $u_{\alpha}=0$. What remains are the even-order moments which we call the backbone moments.
These are various even-order unidirectional moments and various products constructed with their help, up to the triple product of the highest-order even unidirectional moments. For example, for the D3Q125 lattice (the Maxwell lattice generated by the one-dimensional velocity set $V=\{0,\pm1,\pm 3\}$) there are ten different types of the backbone moments arranged in the increasing order from zero to twelve (see Table \ref{table:BBMD3Q125}).
On the other hand, the product lattice can be represented as a collection of shells, each shell contains all the velocities with the same magnitude and symmetry with respect to reflections at the origin and permutation of components. It is important to realize that
\begin{itemize}
\item[]{\it The number of different types of the backbone moments equals the number of shells.}
\end{itemize}
This observation makes it possible to find a one-into-one relation between the backbone moments and the velocity shells which has a form of a two-dimensional key-table (KT).
Let us explain its construction with the example of the D3Q27 product-lattice (see Eq.\ (\ref{eq:D3Q27})).

The backbone moments are then of the four types:
\[1,\  \Pi_{\alpha},\  \Pi_{\alpha}\Pi_{\beta}\ (\alpha\ne\beta),\ \Pi_x\Pi_y\Pi_z.
\]
At the zero-velocity equilibrium $f^{\rm eq}$ (where ${\rm eq}$ is either ${\rm E}$ or ${\rm M}$), these are four different values,
\[
1,\ \Pi_{\alpha}^{\rm eq}=T_0,\   \Pi_{\alpha}^{\rm eq}\Pi_{\beta}^{\rm eq}=T_0^2\ (\alpha\ne\beta),\ \Pi_{x}^{\rm eq}\Pi_{y}^{\rm eq}\Pi_{z}^{\rm eq}=T_0^3.
\]
On the other hand, the D3Q27 lattice is composed of four shells:
\begin{eqnarray*}
V_{0}&=&\{(0,0,0)\},\\
V_{1}&=&\{(\pm 1,0,0),(0,\pm 1,0),(0,0,\pm 1)\},\\
V_{2}&=&\{(\pm 1,\pm 1,0),(\pm 1,0,\pm 1),(0,\pm 1,\pm 1)\},\\
V_{3}&=&\{(\pm 1,\pm 1,\pm 1)\}.
\end{eqnarray*}
The shells $V_{s}$ enumerate the rows in the KT (\ref{eq:D3Q27}). Now, we compute contribution of each shell to each  backbone moment,
introducing the (yet) unknown weights $W_s$ for the velocities of each shell. This corresponds to the $4\times 4$ entries of KT (\ref{eq:D3Q27}). Next, summing up the entries in each of the four columns, and equating the result to the equilibrium value of the corresponding moment, we get
a $4\times4$ linear system for the weights, $W_0+6W_1+12W_2+8W_3=1$, $2W_1+8W_2+8W_3=T_0$, $4W_2+8W_3=T_0^2$, $8W_3=T_0^3$. This system is what remains from (\ref{eq:inverse}) of the D3Q27 at zero-velocity equilibrium. Substituting $T_0=1/3$, we get $W_0=8/27$, $W_1=2/27$, $W_2=1/54$, $W_3=1/216$, the result which we already knew from the product-form.

\begin{equation}
\begin{tabular}{|l|l|l|l|l|l|}
  \hline
 $s$ &  $1$ & $\Pi_{\alpha}$ & $\Pi_{\alpha}\Pi_{\beta}$ ($\alpha\ne\beta$)&$\Pi_x\Pi_y\Pi_z$\\
  \hline
 $0$  &      $W_0$ & $0$ & $0$ &$0$\\
 $1$  &  $6W_1$ & $2W_1$ & $0$ &$0$\\
 $2$  &  $12W_2$ & $8W_2$ & $4W_2$ &$0$\\
 $3$  &  $8W_3$ & $8W_3$ & $8W_3$ &$8W_3$\\
 \hline
\end{tabular}
\label{eq:D3Q27}
\end{equation}

Now, the pruning method with the help of KT (\ref{eq:D3Q27}) consists of erasing one or several rows and of the same number of columns.
Erasing rows is the pruning of the lattice by shell wise discarding of the velocities, whereas erasing rows is sacrificing some of the backbone moments, that is, reducing the accuracy of the UniQuE moment system (\ref{eq:inverse}). Lattices constructed in this way from a Maxwell lattice will be termed sub-Maxwell lattices.
Certainly, in order this procedure to be useful for a further construction of LB models, we should favor lower-order moments as they contain most of the information about the hydrodynamics.
In the present illustrative example of the D3Q27 product-lattice, this means that we should keep the first and the second columns (corresponding to the density and to the diagonal components of the pressure tensor) since these are required for recovering the  Navier-Stokes equations at low Mach numbers, while the higher-order moments (last two columns) can be sacrificed in the pruning procedure.

It is easy to see how the "standard" LB lattices come out as the result of this process.
Erasing the last row ($s=3$) and the last column ($\Pi_x\Pi_y\Pi_z$) in KT (\ref{eq:D3Q27}), summing up the remaining columns, equating the results to the values of the corresponding backbone moments at zero velocity equilibrium at the reference temperature $T_0=1/3$, and solving the resulting $3\times 3$ linear system, gives the weights
$W_0=1/3$, $W_1=1/18$ and $W_2=1/36$ which describe the "standard" D3Q19 lattice. Erasing the third row ($s=2$) and again the last column gives
$W_0=2/9$, $W_1=1/9$ and $W_3=1/72$, which is another standard D3Q15 lattice. Finally, a less standard D3Q13 lattice \blue{\cite{dHumieres01}} corresponds to erasing the second and the last rows ($s=1$ and $s=3$), and two columns, next to the last and the last ($\Pi_{\alpha}\Pi_{\beta}$ and $\Pi_x\Pi_y\Pi_z$), resulting in $W_0=1/2$, $W_2=1/24$.
Note that the present examples illustrates a complete pruning: the three lattices just mentioned are the only pertinent to recovering the isothermal Navier-Stokes equations as the result of pruning of the D2Q27.

Key-tables similar to (\ref{eq:D3Q27}) are obtained in a straightforward manner for product-lattices with any $Q$, and are relatively easy to analyze (for example, for the D3Q125 lattice, the number of types of the backbone moments is an order of magnitude less than the total number of moments, cf. Table \ref{table:BBMD3Q125}).
Following this procedure, we easily identify, for example, the recently introduced D3Q41 lattice \cite{CK09}: The six types of the backbone moments retained are: $1$, $\Pi_{\alpha}$, $\Pi_{\alpha}\Pi_{\beta}$, $R_{\alpha}$, $\Pi_{\alpha}R_{\beta}$ and $\Pi_x\Pi_y\Pi_z$. The retained six shells include the four shells $V_0,\dots,V_3$ of the D3Q27 mentioned above together with $V_4=\{(\pm 3,0,0),(0,\pm 3,0),(0,0,\pm 3)\}$ and
$V_5=\{(\pm 3,\pm 3,\pm 3)\}$. The two latter shells contain 14 velocities which, added to the 27 make up the D3Q41 lattice. Computing the contribution of these six shells to the six backbone moments, and solving the resulting $6\times 6$ linear system, we immediately obtain the corresponding weights,
\begin{align}
\begin{split}
\label{eq:WD3Q41}
W_0 &=1-\frac{1}{81}T_0[270-T_0(263 + 102T_0)],\\
W_1&=\frac{1}{16}T_0[9-T_0(12 + 13T_0)],\\
W_2 &=\frac{1}{2}T_0^3,\\
W_3 &=\frac{1}{64}T_0^2(9-19T_0),\\
W_4 &=\frac{1}{1296}T_0(3T_0-1)(9-T_0),\\
W_5 &=\frac{1}{5184}T_0^3(3T_0-1),
\end{split}
\end{align}
which are positive at $T_0=1-\sqrt{2/5}$ (see Table \ref{table:minimal_lattices}), and coincide with those reported in \cite{CK09}.

The pruning of the Maxwell lattice using its KT derives the important information, the weights $W_s$ corresponding to the retained shells (for the pruned lattices, the weights are not products of any one-dimensional weights any longer, as it was for the product-lattice). This immediately triggers the option of equilibration by minimizing the corresponding entropy \cite{CK09}.
The equilibration is performed under fixed density and velocity, which are now defined on the sub-Maxwell lattice.

Finally, we remark that KT establishes the most "fine-grained" (one-into-one) correspondence between (groups of ) velocities and moments (it is not possible to establish a "finer" correspondence between the moments and the velocities than that provided by KT since many velocities contribute to each particular moment). The relation between velocity shells and backbone moments, as presented by KT, is therefore the optimal setting for pruning, in general.

\subsection{Projection pruning}
\label{sec:PP}

The advantage of the above entropy pruning (EP) is that, once the weights are found from KT, we do not need to care about the higher-order moments since their equilibrium values will be decided by the corresponding equilibrium $f^{\rm E}$.
The disadvantage is that we (still) need to solve a nonlinear minimization problem to find $f^{\rm E}$. Therefore, a different way of pruning can be offered which avoids the entropy minimization and is much easier to execute.

This route is, in fact, a continuation of the KT to include a part of the moment system (\ref{eq:inverse}), addressing also the moments which were washed out at the zero-velocity equilibrium. Let us again explain it with the example of D3Q27 (see Table \ref{table:PP}).
First, we group all the moments (\ref{eq:inverse}) according to their (usual) order from $0$ to $6$ (the highest-order moment corresponds to the triple product $\Pi_x\Pi_y\Pi_z$), writing the backbone moments first (first row of Table \ref{table:PP}). For each lattice found from the above analysis of KT, we fill out the corresponding row by retaining (a part of) the moments (\ref{eq:inverse}), moving from the left to the right (from the lower to higher order moments). For example, for the D3Q19 lattice (second row in Table \ref{table:PP}), we first include all the moments in the columns $0$, $1$ and $2$ as they define the basic fields (density and velocity), and the pressure tensor. In the column $3$, we can include all the third-order moments except for $u_xu_yu_z$ because $M_{111}$ degenerates on the shells retained in the D3Q19: Since any velocity vector of D3Q19 contains at least one zero component, we have $v_{(i)}v_{(j)}v_{(k)}=0$ ($i\ne j\ne k$) for any vector. This degeneracy precludes the moment $M^*_{111}=u_xu_yu_z$ to be retained by the moment system of D3Q19, and we proceed to the next column, where we can retain only the three backbone moments.
In the case of $D3Q15$, the situation is opposite at the column $3$: while the moment $M_{111}$ is non-degenerate, and thus the value $M_{111}^*$ can be now retained,
the three pairs of moments, $M_{120}$ and $M_{102}$, $M_{210}$ and $M_{012}$, and $M_{021}$ and $M_{201}$ become degenerated, and only the three linearly independent combinations can be retained. For that, we choose symmetric combinations, as shown in Table \ref{table:PP}. Finally, the three backbone moments are degenerated by D3Q15, $M_{220}=M_{202}=M_{022}$, and we are able to retain their symmetric combination.
Similar considerations apply also for the last (D3Q13) lattice reported in Table \ref{table:PP}.

\begin{table}
\begin{tabular}{|l|l|l|l|l|l|l|l|}
  \hline
            &  $0$ &   $1$      & $2$ &  $3$  & $4$ &   $5$         &    $6$\\
\hline
Backbone    &  $1$ &   & $\Pi_{x},\Pi_y,\Pi_z$ &  & $\Pi_x\Pi_y,\Pi_y\Pi_z,\Pi_x\Pi_z$ &            &    $\Pi_x\Pi_y\Pi_z$\\
            &      & $u_x,u_y,u_z$ & $u_{x}u_{y},u_yu_z,u_xu_z$ & $u_xu_yu_z$ &   $u_xu_y\Pi_z$  &   $u_x\Pi_y\Pi_z$&             \\
{\rm D3Q27} &      &               &                            & $u_{x}\Pi_{y}, u_x\Pi_z$ & $u_xu_z\Pi_y$& $u_y\Pi_x\Pi_z$ &\\
            &      &               &                            & $u_{y}\Pi_{x}, u_y\Pi_z$ & $u_yu_z\Pi_x$ & $u_z\Pi_x\Pi_y$&\\
            &      &               &                            & $u_{z}\Pi_{x}, u_z\Pi_y$ &                 & &\\
\hline
Backbone    &  $1$ &   & $\Pi_{x},\Pi_y,\Pi_z$ &  & $\Pi_x\Pi_y,\Pi_y\Pi_z,\Pi_x\Pi_z$ &            &    \\
            &      & $u_x,u_y,u_z$ & $u_{x}u_{y},u_yu_z,u_xu_z$ &  &    &   &             \\
D3Q19       &      &               &                            & $u_{x}\Pi_{y}, u_x\Pi_z$ & & &\\
            &      &               &                            & $u_{y}\Pi_{x}, u_y\Pi_z$ & & &\\
            &      &               &                            & $u_{z}\Pi_{x}, u_z\Pi_y$ & & &\\

 \hline
Backbone    &  $1$ &   & $\Pi_{x},\Pi_y,\Pi_z$ &  & $\Pi_x\Pi_y+\Pi_y\Pi_z+\Pi_x\Pi_z$ &            &    \\
            &      & $u_x,u_y,u_z$ & $u_{x}u_{y},u_yu_z,u_xu_z$ & $u_xu_yu_z$ &    &   &             \\
D3Q15       &      &               &                            & $u_{x}(\Pi_{y}+\Pi_z)$ & & &\\
            &      &               &                            & $u_{y}(\Pi_{x}+\Pi_z)$ & & &\\
            &      &               &                            & $u_{z}(\Pi_{x}+\Pi_y)$ & & &\\

 \hline
Backbone    &  $1$ &   & $\Pi_{x},\Pi_y,\Pi_z$ &  &  &            &    \\
            &      & $u_x,u_y,u_z$ & $u_{x}u_{y},u_yu_z,u_xu_z$ &  &    &   &             \\
D3Q13       &      &               &                            & $u_{x}(\Pi_{y}-\Pi_z)$ & & &\\
            &      &               &                            & $u_{y}(\Pi_{x}-\Pi_z)$ & & &\\
            &      &               &                            & $u_{z}(\Pi_{x}-\Pi_y)$ & & &\\

 \hline
\end{tabular}
\caption{Projection pruning of the D3Q27 UniQuE moment system (\ref{eq:inverse}). Moments are grouped in columns, according to their order, from $0$ to $6$.
Each row contains the moments retained by a particular lattice. Backbone moments are indicated first. First row (D3Q27) represents the full UniQuE moment system (\ref{eq:inverse}). Filling out the rows corresponding to D3Q19, D3Q15 and D3Q13 is explained in the text. Maxwellization (construction of the equilibrium) is achieved by replacing $\Pi_{\alpha}\to\Pi_{\alpha}^{\rm M}$, where $\Pi_{\alpha}^{\rm M}=T_0+u_{\alpha}^2$, and $T_0=1/3$ is the reference temperature. Example of D3Q19 is presented in Appendix \ref{appendix:D3Q27}, Eqs.\ (\ref{eq:D3Q19}) and (\ref{eq:MD3Q19}).
\blue{Example of D3Q13 is further discussed in Appendix \ref{appendix:D3Q13}.}}
\label{table:PP}
\end{table}

Now, the number of retained moments in each row of Table \ref{table:PP} equals to the number of the populations of the corresponding lattice.
Consequently, these moment relations, with the right hand side given by Table \ref{table:PP}, can be readily inverted to derive an analog of the UniQuE,
\begin{equation}
\label{eq:UniQuEPP}
{f}^*_{(i,j,k)}=\rho{\varphi}_{(i,j,k)}({\cal M}_x,{\cal M}_y,{\cal M}_z),
\end{equation}
where now $(i,j,k)$ spans not the whole range of indices but only those corresponding to the retained shells. Consequently, ${\varphi}_{(i,j,k)}$ do not have the form of a product of the unidirectional functions (\ref{eq:UniQuE}) (although it resembles the latter, as illustrated by the D3Q19, see Eq.\ (\ref{eq:D3Q19}) in Appendix \ref{appendix:D3Q27}).
Function ${f}^*$ (\ref{eq:UniQuEPP}) represents the UniQuE moment system (\ref{eq:inverse}) in the best possible way allowed by the reduced number of velocities, thereby providing a projection of the
D3Q27 lattice onto the corresponding pruned lattice.
For that reason, we term the present method as projection pruning (PP), in order to distinguish it from the entropy pruning.

Since PP derives ${f}^*$ (\ref{eq:UniQuEPP}) from the moment system of the Maxwell lattice (\ref{eq:inverse}), the notion of the equilibrium for it is also a derivative of the corresponding results for the UniQuE (\ref{eq:UniQuE}): It is either equilibration, induced by the equilibrium values ${\cal M}^{\rm E}_{\alpha}=\{u_{\alpha},{\cal N}^{\rm E}_\alpha\}$ of the corresponding one-dimensional Maxwell lattice,
\begin{equation}
\label{eq:EPP}
{f}_{(i,j,k)}^{\rm E}=\rho{\varphi}_{(i,j,k)}\left(\{u_x, {\cal N}^{\rm E}_x\}, \{u_y, {\cal N}^{\rm E}_y\}, \{u_z, {\cal N}^{\rm E}_z\}\right),
\end{equation}
or Maxwellization, induced by the Maxwell values of the same one-dimensional moments ${\cal M}^{\rm M}_{\alpha}=\{u_{\alpha},{\cal N}^{\rm M}_\alpha\}$
\begin{equation}
\label{eq:MPP}
{f}_{(i,j,k)}^{\rm M}=\rho{\varphi}_{(i,j,k)}\left(\{u_x, {\cal N}^{\rm M}_x\}, \{u_y, {\cal N}^{\rm M}_y\}, \{u_z, {\cal N}^{\rm M}_z\}\right).
\end{equation}
In Appendix \ref{appendix:D3Q27}, we give example of $f^*$ (\ref{eq:UniQuEPP}) and $f^{\rm M}$ (\ref{eq:MPP}) for the D3Q19 sub-Maxwell lattice (Eqs.\ (\ref{eq:D3Q19}) and (\ref{eq:MD3Q19}), respectively). All these considerations are readily applicable to the projection pruning of any product-lattice.

\blue{Finally, we note that, as the result of the present complete pruning, we arrive at the set of admissible lattices and corresponding quasi-equilibria and equilibria. The question of which LB model can be supported by a particular sub-Maxwell lattice remains beyond the scope of this analysis. However, this is easily done upon studying the set of moments retained after the pruning. Note that, in general, the familiar single relaxation time lattice BGK model may be not sufficient, and more general kinetic models need to be addressed, such as the quasi-equilibrium models \cite{Gorban94,Levermore96,Ansumali07,Asinari09} which make use of the quasi-equilibrium along with the equilibrium, or the multiple relaxation times (MRT) models (see, e.\ g., a paper by I. Ginzburg \cite{Ginzburg08} and references therein). As an illustration, a two-step quasi-equilibrium model for incompressible flow is derived for the D3Q13 lattice in Appendix \ref{appendix:D3Q13}, utilizing the above UniQuE quasi-equilibrium (\ref{eq:UniQuEPP}).}

\section{Discussion}
\label{sec:conclusion}

Maxwell's derivation of the equilibrium distribution function in a gas predated Boltzmann's fundamental $H$-theorem and the specification of the equilibrium as the minimum of $H$. Maxwell's argument was based on the independence of the equilibrium on the direction, resulting from the multiplication of the unidirectional equilibrium functions. Both approaches, Maxwell's and Boltzmann's, result in the same Gaussian equilibrium.

In this paper, we followed closely the Maxwell's path, exploiting the symmetry of the product for the purpose of constructing  LB models. The main result of the present theory is the constructive approach to better, Galilean invariant higher-order LB models.
Here we summarize the construction of LB developed above, and make further comments on these findings.
\begin{itemize}

\item Construction of  any three-dimensional LB takes it origin  in one dimension. For a given lattice, we consider the closure relation and
derive the reference temperature via the matching condition. The reference temperature does not change in any further step of the construction. It is the characteristics of the one-dimensional lattice, and of all the lattices induced by the one-dimensional lattice in three dimensions (Maxwell and sub-Maxwell lattices).

\item Let us give another interpretation of the matching condition. The Maxwell moments arise from the Gaussian distribution (\ref{eq:M}). That means, they obey a recurrence relation which expresses the higher-order moments in terms of the two lower moments (the mean and the variance). This recurrence relation is well known and is not reproduced here. Important is that the moments of the Gaussian prolong: Once the first and the second moments are known, the rest of the moments are computed from the recurrence relation. Now, with a finite number of velocities $Q$ (odd), we can reproduce first $Q$ moments of the Gaussian (including normalization). However, this does not say anything yet whether or not the moment sequence will be prolonged. Because of the closure relation, such a prolongation is restricted  to $M_{(Q)}$ and $M_{(Q+1)}$, the former is odd and was used in the matching condition, the latter is even and leads to the same matching condition. Thus, the matching condition verifies a restricted Gaussian prolongation, it checks the moments which are not independent of the first $Q$ moments (by the closure relation). But higher-order moments of the Gaussian are also dependent on the lower-order moments (through recurrence relation). So, the matching condition seeks consistence between the two different relations, the one is the closure relation (pertinent to the discreteness of the velocities), and the other pertinent to the Gaussian. This verification of the restricted prolongation is thus the verification of the restricted Gaussian feature for the given velocity set, and it reduces to the verification of the reference temperature, as it was done in sec.\ \ref{sec:D1}. In other words, important is not the reproducing of the $Q$ moments of the Gaussian with $Q$ populations (this can be done by any velocity set) but rather the prolongation property, which is the matching condition.

\item Transition to three dimensions begins with the construction of the Maxwell lattice, for which one defines UniQuE, the special quasi-equilibrium in the form of a product of unidirectional functions. UniQuE has  remarkably simple moment relations (products of unidirectional moments), and reduces the analysis of the moment systems from $Q^3$ to $3Q-2$ dimensions.
 Construction of the equilibrium on the Maxwell lattice requires only the unidirectional information.

\item All other lattices are obtained as a pruning of the product lattice. The method of key-table reduces the problem of constructing the entropy function of the pruned lattice to analyzing a two-dimensional table and verifying consistency and solving linear problems. For large $Q$, this can be achieved with standard tools of linear programming (verification of consistency of linear systems). However, even the intuitive search for good sub-Maxwell lattices is possible with the key-table thanks to its relative simplicity.

\item Finally, the projection pruning is introduced as an extension of the key-table, which enables to derive UniQuE and Maxwellization for pruned lattices. This requires only solving linear systems. Maxwellization on the pruned lattices is a promising approach to higher-order lattices due to a relative simplicity of construction.

\item Derivation of any lattice in any dimension begins with finding the reference temperature and the equilibrium at zero velocity (weights). After that, there are two options to continue, equilibration or Maxwellization. The strong point about equilibration is that it is based on the entropy minimization, and stability theorems (Boltzmann's $H$-theorems) can be proved in that case for various LB realizations. However, in order to obtain the equilibrium on that route, one needs to solve a nonlinear minimization problem which, in most cases, can be only done within an approximation. On the other hand, in the Maxwellization approach, the corresponding equilibrium is constructed much easier, even for sub-Maxwell lattices it requires only solving linear systems.
\item We note that specific cases of UniQuE were used recently in order to construct quasi-equilibrium LB models with enhanced stability \cite{Asinari09,Asinari09a}, and to enhance Galilean invariance of LB models on standard lattices \cite{Prasianakis09}.
\item
Finally, we point out that
UniQuE represents an exact and systematic alternative to
other closure procedures reported in literature, not necessarily in the LB context. For
example, a moment-inversion algorithm was developed
recently based on Cholesky decomposition of the velocity
covariance matrix and repeated application of
one-dimensional quadrature for dilute gas-particle flows
\cite{Fox1}. Even though it is well recognized that the
moment-inversion problem admits exact solution in one
dimension (e.g. by product-difference algorithm),
defining the linear system used to solve for the weights
and the abscissas in multi-dimensional case is still an
open question. In particular, it is recognized \cite{Fox1} that
the most suitable algorithms (Cholesky decomposition,
method of eigenvectors etc) appears as problem dependent. In
this context, UniQuE offers a simple and general framework
to develop closure models starting from the analytical one-dimensional solution. Fixed abscissas used by Maxwell
lattices do not represent a limit, since the closure
relations in terms of the considered moments can be
derived explicitly and implemented in functional form in
the generalized hydrodynamic equations. Outcomes of this
procedure are expected for granular flows,
polydisperse liquid sprays undergoing droplet coalescence
and evaporation and, more generally, aerosol
dynamics \cite{Fox1}. These problems will be addressed in our future work.
\end{itemize}
I.V.K. gratefully acknowledges support of CCEM-CH, and thanks E. Chiavazzo for a help with some algebra.


\begin{appendix}

\section{How to find closure relation and verify reference temperature for a given velocity set}
\label{appendix:generalQ}
For $Q$ velocities ($Q$ odd), one writes $v_{(i)}^Q=a_{Q-2}v_{(i)}^{Q-2}+a_{(Q-4)}v_{(i)}^{Q-4}+\dots+a_{1}v_{(i)}$, substitutes $(Q-1)/2$ different non-zero values for the velocities and solves the linear system for the coefficients $a_{Q-2},\dots,a_{1}$. Once the latter are obtained, we use $M_{(n)}^{\rm M}=b_nT_{0}^{(n-1)/2}u+O(u^3)$ ($n$ odd) with $b_n=1\times 3\times 5\dots\times n$. Matching condition of linear in $u$ terms results in the algebraic equation for
the reference temperature, $b_{Q}T_0^{(Q-1)/2}-a_{Q-2}b_{Q-2}T_0^{(Q-3)/2}-a_{(Q-4)}b_{Q-4}T_0^{(Q-5)/2}-\dots-a_{1}=0$. Positive roots (if they exist) define the reference temperature. If no positive roots are available, the corresponding lattice is ruled out of a further consideration.

\section{Maxwell lattices and roots of Hermite polynomials}
\label{appendix:D1Q5}
In Ref.\ \cite{CK_PRL2}, it was argued that one-dimensional Maxwell lattices have ratios of the velocities that approximate the ratios of the roots of Hermite polynomials. We recover this argument here from the closure relation and the matching condition, considering the example of D1Q5.
Without loss of generality, the one-dimensional velocities are set as $V=\{0,\pm 1,\pm r\}$, where $r>1$.
The closure relation then reads: $v_{(i)}^5=(1+r^2)v_{(i)}^3-r^2v_{(i)}$. The matching condition results in the following quadratic equation
for the reference temperature: $15T_0^2-3(1+r^2)T_0+r^2=0$. This equation has positive real-valued solutions if $r\ge r^*$, where $r^*=\sqrt{t^*}$ with $t^*$ the larger root of another quadratic equation, $3(1+t)^2-20t=0$. From the latter we find $t^*=(7+2\sqrt{10})/3$, and taking the root of it, we find $r^*=(\sqrt{5}+\sqrt{2})/\sqrt{3}$. This is nothing but the ratio between the two non-trivial roots of the $5$-th order Hermite polynomial (these roots are $\{0,\pm\sqrt{5-\sqrt{10}},\pm\sqrt{5+\sqrt{10}}\}$). Thus, we have recovered the argument of Ref.\ \cite{CK_PRL2} by a different
consideration.

\section{Main theorem about UniQuE}
\label{appendix:theoremUniQuE}
We here give the proof of the theorem of sec.\ \ref{sec:UniQuE} which characterizes the UniQuE population (\ref{eq:UniQuE}) as the quasi-equilibrium. We restore to expanded notation:
For $D=3$, the density is defined as
\begin{align}
\begin{split}
\label{eq:2constraints1}
\rho&=\sum_{i\in V}\sum_{j\in V}\sum_{k\in V}f_{(i,j,k)},
\end{split}
\end{align}
while the unidirectional moments are
\begin{align}
\begin{split}
\label{eq:2constraints2}
\rho M_x^{(n)}&=\sum_{i\in V}\sum_{j\in V}\sum_{k\in V}v_{(i)}^nf_{(i,j,k)},\ n=1,\dots,Q-1,\\
\rho M_y^{(n)}&=\sum_{i\in V}\sum_{j\in V}\sum_{k\in V}v_{(j)}^nf_{(i,j,k)},\ n=1,\dots,Q-1,\\
\rho M_z^{(n)}&=\sum_{i\in V}\sum_{j\in V}\sum_{k\in V}v_{(k)}^nf_{(i,j,k)},\ n=1,\dots,Q-1,\\
\end{split}
\end{align}
The moment densities $M_{\alpha}^{(n)}$, $\alpha=x,y,z$ are termed unidirectional in order to reflect the fact that only the $x$-component $v_{(i)}$ of the three-dimensional velocity vector $v_{(i,j,k)}=(v_{(i)},v_{(j)},v_{(k)})$ participates in the definition of  $M_{x}^{(n)}$, while only the $y$-component $v_{(j)}$ participates in the definition of $M_{y}^{(n)}$, etc.
Finally, we denote
\begin{align}
\begin{split}
{\cal M}_x&=\left\{M_x^{(1)},\dots,M_x^{(Q-1)}\right\},\\
{\cal M}_y&=\left\{M_y^{(1)},\dots,M_y^{(Q-1)}\right\},\\
{\cal M}_z&=\left\{M_z^{(1)},\dots,M_z^{(Q-1)}\right\}.\\
\end{split}
\end{align}

\noindent {\bf Theorem}: Let the parameters ${\cal M}_{\alpha}$ take their values in the positivity domain, ${\cal M}_{\alpha}\in \Omega$, $\alpha=x,y,z$. Then the minimizer of the entropy function $H$,

\begin{equation}\label{eq:HwithW}
H=\sum_{i\in V}\sum_{j\in V}\sum_{k\in V}f_{(i,j,k)}\ln\left(\frac{f_{(i,j,k)}}{W_{(i)}W_{(j)}W_{(k)}}\right),
\end{equation}
under the constraints (\ref{eq:2constraints1}) and (\ref{eq:2constraints2}) is given by the product-function  (\ref{eq:UniQuE}).
\bigskip

\noindent{\bf Proof:}
The extremum condition is written
\begin{equation}
\ln\left(\frac{f^*_{(i,j,k)}}{W_{(i)}W_{(j)}W_{(k)}}\right)=\Lambda -1 +\sum_{n=1}^{Q-1}\lambda_{x}^{(n)}v_{(i)}^n
+\sum_{n=1}^{Q-1}\lambda_{y}^{(n)}v_{(j)}^n
+\sum_{n=1}^{Q-1}\lambda_{z}^{(n)}v_{(k)}^n,
\end{equation}
where $\Lambda$ is the Lagrange multiplier corresponding to the density constraint (\ref{eq:2constraints1}), and $\lambda_{\alpha}^{(n)}$ are the Lagrange multipliers corresponding to the unidirectional moment constraints (\ref{eq:2constraints2}).
This can be rewritten as
\begin{equation}\label{eq:Fstar}
f^*_{(i,j,k)}=\rho X_{(i)}Y_{(j)}Z_{(k)},
\end{equation}
with
\begin{align}
\begin{split}\label{eq:Psistar}
X_{(i)}=W_{(i)}\exp\left(\frac{\Lambda-1-\ln\rho}{3}+\sum_{n=1}^{Q-1}\lambda_{x}^{(n)}v_{(i)}^n\right),\\
Y_{(j)}=W_{(j)}\exp\left(\frac{\Lambda-1-\ln\rho}{3}+\sum_{n=1}^{Q-1}\lambda_{y}^{(n)}v_{(j)}^n\right),\\
Z_{(k)}=W_{(k)}\exp\left(\frac{\Lambda-1-\ln\rho}{3}+\sum_{n=1}^{Q-1}\lambda_{z}^{(n)}v_{(k)}^n\right),
%
\end{split}
\end{align}
Substituting (\ref{eq:Fstar})
into the constraints (\ref{eq:2constraints1}) and (\ref{eq:2constraints2}), the latter becomes

\begin{eqnarray}
\left(\sum_{i\in V}X_{(i)}\right)\left(\sum_{j\in V}Y_{(j)}\right)\left(\sum_{k\in V}Z_{(k)}\right)=1,\label{eq:density}\\
\left(\sum_{i\in V}v^n_{(i)}X_{(i)}\right)\left( \sum_{j\in V}Y_{(j)}\right)\left(\sum_{k\in V}Z_{(k)}\right)=  M_{x}^{(n)},\ n=1,\dots, Q-1,\label{eq:Xrest}\\
\left(\sum_{j\in V}v^n_{(j)}Y_{(i)}\right)\left( \sum_{i\in V}X_{(i)}\right)\left(\sum_{k\in V}Z_{(k)}\right)= M_{y}^{(n)},\ n=1,\dots, Q-1,\label{eq:Yrest}\\
\left(\sum_{k\in V}v^n_{(k)}Z_{(k)}\right)\left( \sum_{i\in V}X_{(i)}\right)\left(\sum_{j\in V}Y_{(k)}\right)= M_{z}^{(n)},\ n=1,\dots, Q-1.\label{eq:Zrest}\\
\end{eqnarray}
Equation (\ref{eq:density}) admits a solution (the normalization condition),
\begin{equation}
\label{Lambda}
 \sum_{i\in V} X_{(i)} =1,\ \sum_{j\in V}Y_{(j)}=1,\ \sum_{k\in V}Z_{(k)}=1,
\end{equation}
which implies for the rest of the conditions, Eqs.\ (\ref{eq:Xrest}), (\ref{eq:Yrest}) and (\ref{eq:Zrest}),
\begin{align}
\begin{split}
\label{eq:decomposedX}
&\sum_{i\in V} X_{(i)}=1,\\
&\sum_{i\in V} v^n_{(i)}X_{(i)}=M_{x}^{(n)},\ n=1,\dots, Q-1,\\
\end{split}
\end{align}
\begin{align}
\begin{split}
\label{eq:decomposedY}
&\sum_{j\in V}Y_{(j)}=1,\\
&\sum_{j\in V} v^n_{(j)}Y_{(j)}=M_{y}^{(n)},\ n=1,\dots, Q-1,\\
\end{split}
\end{align}
\begin{align}
\begin{split}
\label{eq:decomposedZ}
&\sum_{k\in V}Z_{(k)}=1,\\
&\sum_{k\in V} v^n_{(k)}Z_{(j)}=M_{z}^{(n)},\ n=1,\dots, Q-1.\\
\end{split}
\end{align}
Now, each of the  problems (\ref{eq:decomposedX}), (\ref{eq:decomposedY}) and (\ref{eq:decomposedZ}) is equivalent to the one-dimensional problem solved in sec.\ \ref{sec:D1} and which defines the one-dimensional functions $\varphi_{(i)}({\cal M})$, and thus the solution of each of these problems separately is given by the unidirectional quasi-equilibrium, viz.
\begin{align}
\begin{split}
\label{eq:Psia}
X_{(i)}&=\varphi_{(i)}(M_{x}^{(1)},\dots, M_{x}^{(Q-1)}),\\
Y_{(j)}&=\varphi_{(j)}(M_{y}^{(1)},\dots, M_{y}^{(Q-1)}),\\
Z_{(k)}&=\varphi_{(k)}(M_{z}^{(1)},\dots, M_{z}^{(Q-1)}).\\
\end{split}
\end{align}
With (\ref{eq:Psia}) and (\ref{Lambda}), we find a solution in the form (\ref{eq:UniQuE}). The proof is completed by reminding that the minimum of a convex function under a set of linear constraints is unique.


\section{D3Q27 and D3Q19: UniQuE, equilibration and Maxwellization}
\label{appendix:D3Q27}
Here we collect various populations for the D3Q27 Maxwell lattice and for the D3Q19 sub-Maxwell lattice mentioned in the paper.
The list begins with the UniQuE (\ref{eq:UniQuE}) for the D3Q27:
\begin{align}
\begin{split}\label{eq:UniQuED3Q27}
f^*_{(0,0,0)}&=\rho(1-\Pi_x)(1-\Pi_y)(1-\Pi_z),\\
f^*_{(\pm1,0,0)}&=\frac{1}{2}(\Pi_x\pm u_x)\rho(1-\Pi_y)(1-\Pi_z),\\
f^*_{(0,\pm1,0)}&=\frac{1}{2}\rho(1-\Pi_x)(\Pi_y\pm u_y)(1-\Pi_z),\\
f^*_{(0,0,\pm1)}&=\frac{1}{2}\rho(1-\Pi_x)(1-\Pi_y)(\Pi_z\pm u_z),\\
f^*_{(\pm1,\pm1,0)}&=\frac{1}{4}\rho(\Pi_x\pm u_x)(\Pi_y\pm u_y)(1-\Pi_z),\\
f^*_{(0,\pm1,\pm1)}&=\frac{1}{4}\rho(1-\Pi_x)(\Pi_y\pm u_y)(\Pi_z\pm u_z),\\
f^*_{(\pm1,0\pm1)}&=\frac{1}{4}\rho(\Pi_x\pm u_x)(1-\Pi_y)(\Pi_z\pm u_z),\\
f^*_{(\pm1,\pm1\pm1)}&=\frac{1}{8}\rho(\Pi_x\pm u_x)(\Pi_y\pm u_y)(\Pi_z\pm u_z).\\
\end{split}
\end{align}
Note that, when setting $\Pi_z=0$ in the nine populations, $f^*_{(0,0,0)}$, $f^*_{(\pm1,0,0)}$, $f^*_{(0,\pm1,0)}$, and $f^*_{(\pm1,\pm1,0)}$ (\ref{eq:UniQuED3Q27}), we obtain the UniQuE on the two-dimensional D2Q9 lattice:
\begin{align}
\begin{split}\label{eq:UniQuED2Q9}
f^*_{(0,0)}&=\rho(1-\Pi_x)(1-\Pi_y),\\
f^*_{(\pm1,0)}&=\frac{1}{2}(\Pi_x\pm u_x)\rho(1-\Pi_y),\\
f^*_{(0,\pm1)}&=\frac{1}{2}\rho(1-\Pi_x)(\Pi_y\pm u_y),\\
f^*_{(\pm1,\pm1)}&=\frac{1}{4}\rho(\Pi_x\pm u_x)(\Pi_y\pm u_y).\\
\end{split}
\end{align}
This two-dimensional UniQuE was used in \cite{Asinari09} for a construction of a class of two relaxation times models with enhanced stability.

Equilibration of (\ref{eq:UniQuED3Q27}) is achieved upon substituting the equilibrium one-dimensional pressure,
\begin{equation}
\Pi^{\rm E}_{\alpha}=\frac{1}{3}\left(2\sqrt{1+3u_{\alpha}^2}-1\right),
\end{equation}
into (\ref{eq:UniQuED3Q27}):
\begin{align}
\begin{split}\label{eq:ED3Q27eq}
f^{\rm E}_{(0,0,0)}&=\frac{8}{27}\rho\left(2-\sqrt{1+3u_{x}^2}\right)\left(2-\sqrt{1+3u_{y}^2}\right)\left(2-\sqrt{1+3u_{z}^2}\right),\\
f^{\rm E}_{(\pm1,0,0)}&=\frac{2}{27}\rho\left(2\sqrt{1+3u_{x}^2}-1\pm 3u_x\right)\left(2-\sqrt{1+3u_{y}^2}\right)\left(2-\sqrt{1+3u_{z}^2}\right),\\
f^{\rm E}_{(0,\pm1,0)}&=\frac{2}{27}\rho\left(2-\sqrt{1+3u_{x}^2}\right)\left(2\sqrt{1+3u_{y}^2}-1\pm 3u_y\right)\left(2-\sqrt{1+3u_{z}^2}\right),\\
f^{\rm E}_{(0,0,\pm1)}&=\frac{2}{27}\rho\left(2-\sqrt{1+3u_{x}^2}\right)\left(2-\sqrt{1+3u_{y}^2}\right)\left(2\sqrt{1+3u_{z}^2}-1\pm 3u_z\right),\\
f^{\rm E}_{(\pm1,\pm1,0)}&=\frac{1}{54}\rho\left(2\sqrt{1+3u_{x}^2}-1\pm 3u_x\right)\left(2\sqrt{1+3u_{y}^2}-1\pm 3u_y\right)\left(2-\sqrt{1+3u_{z}^2}\right),\\
f^{\rm E}_{(0,\pm1,\pm1)}&=\frac{1}{54}\rho\left(2-\sqrt{1+3u_{x}^2}\right)\left(2\sqrt{1+3u_{y}^2}-1\pm 3u_y\right)\left(2\sqrt{1+3u_{z}^2}-1\pm 3u_z\right),\\
f^{\rm E}_{(\pm1,0\pm1)}&=\frac{1}{54}\rho\left(2\sqrt{1+3u_{x}^2}-1\pm 3u_x\right)\left(2-\sqrt{1+3u_{y}^2}\right)\left(2\sqrt{1+3u_{z}^2}-1\pm 3u_z\right),\\
f^{\rm E}_{(\pm1,\pm1\pm1)}&=\frac{1}{216}\rho\left(2\sqrt{1+3u_{x}^2}-1\pm 3u_x\right)\left(2\sqrt{1+3u_{y}^2}-1\pm 3u_y\right)\left(2\sqrt{1+3u_{z}^2}-1\pm 3u_z\right).\\
\end{split}
\end{align}
Weights $W_{s}$, corresponding to various shells $s=0,1,2,3$ (see (\ref{eq:D3Q27})), are numerical pre-factors in these expressions.
Equilibrium (\ref{eq:ED3Q27eq}) was derived in \cite{Ansumali03} by a direct minimization of entropy in three dimensions.
Positivity domain of (\ref{eq:ED3Q27eq}) (all populations are non-negative simultaneously) is a box with the edge $2$ centered at the origin of the three-dimensional parameter space $(u_x,u_y,u_z)$: $\Omega_{\rm D3Q27}^{\rm E}=\{\bm{u}: |u_\alpha|\le 1, \alpha=x,y,z\}$.

Maxwellization of (\ref{eq:UniQuED3Q27}) is found upon a substitution into (\ref{eq:UniQuED3Q27}) the Maxwell expression for diagonal components of the pressure tensor at unit density,
\begin{equation}
\label{eq:MaxwellizatorD1Q3}
\Pi^{\rm M}_{\alpha}=\frac{1}{3}\left(1+3u_{\alpha}^2\right),
\end{equation}
which gives explicitly
\begin{align}
\begin{split}\label{eq:MD3Q27eq}
f^{\rm M}_{(0,0,0)}&=\frac{8}{27}\rho\left(1-\frac{3}{2}u_{x}^2\right)\left(1-\frac{3}{2}u_{y}^2\right)\left(1-\frac{3}{2}u_{z}^2\right),\\
f^{\rm M}_{(\pm1,0,0)}&=\frac{2}{27}\rho\left(1\pm 3u_x+3u_{x}^2\right)\left(1-\frac{3}{2}u_{y}^2\right)\left(1-\frac{3}{2}u_{z}^2\right),\\
f^{\rm M}_{(0,\pm1,0)}&=\frac{2}{27}\rho\left(1-\frac{3}{2}u_{x}^2\right)\left(1\pm 3u_y+3u_{y}^2\right)\left(1-\frac{3}{2}u_{z}^2\right),\\
f^{\rm M}_{(0,0,\pm1)}&=\frac{2}{27}\rho\left(1-\frac{3}{2}u_{x}^2\right)\left(1-\frac{3}{2}u_{y}^2\right)\left(1\pm 3u_z+3u_{z}^2\right),\\
f^{\rm M}_{(\pm1,\pm1,0)}&=\frac{1}{54}\rho\left(1\pm 3u_x+3u_{x}^2\right)\left(1\pm 3u_y+3u_{y}^2\right)\left(1-\frac{3}{2}u_{z}^2\right),\\
f^{\rm M}_{(0,\pm1,\pm1)}&=\frac{1}{54}\rho\left(1-\frac{3}{2}u_{x}^2\right)\left(1\pm 3u_y+3u_{y}^2\right)\left(1\pm 3u_z+3u_{z}^2\right),\\
f^{\rm M}_{(\pm1,0\pm1)}&=\frac{1}{54}\rho\left(1\pm 3u_x+3u_{x}^2\right)\left(1-\frac{3}{2}u_{y}^2\right)\left(1\pm 3u_z+3u_{z}^2\right),\\
f^{\rm M}_{(\pm1,\pm1\pm1)}&=\frac{1}{216}\rho\left(1\pm 3u_x+3u_{x}^2\right)\left(1\pm 3u_y+3u_{y}^2\right)\left(1\pm 3u_z+3u_{z}^2\right).\\
\end{split}
\end{align}
Positivity domain of (\ref{eq:MD3Q27eq}) is the box with the edge $2\sqrt{2/3}$: $\Omega_{\rm D3Q27}^{\rm M}=\{\bm{u}: |u_\alpha|\le \sqrt{2/3}, \alpha=x,y,z\}$.

For the D3Q19 sub-Maxwell lattice, the analog of UniQuE constructed by projection pruning (\ref{eq:UniQuEPP}) is:
\begin{align}
\begin{split}\label{eq:D3Q19}
{f}^*_{(0,0,0)}&=\rho(1-\Pi_x-\Pi_y-\Pi_z+\Pi_x\Pi_y+\Pi_y\Pi_z+\Pi_x\Pi_z),\\
{f}^*_{(\pm1,0,0)}&=\frac{1}{2}\rho(1-\Pi_y-\Pi_z)(\Pi_x\pm u_x),\\
{f}^*_{(0,\pm1,0)}&=\frac{1}{2}\rho(1-\Pi_x-\Pi_z)(\Pi_y\pm u_y),\\
{f}^*_{(0,0,\pm1)}&=\frac{1}{2}\rho(1-\Pi_x-\Pi_y)(\Pi_z\pm u_z),\\
{f}^*_{(\pm1,\pm1,0)}&=\frac{1}{4}\rho(\Pi_x\pm u_x)(\Pi_y\pm u_y),\\
{f}^*_{(0,\pm1,\pm1)}&=\frac{1}{4}\rho(\Pi_y\pm u_y)(\Pi_z\pm u_z),\\
{f}^*_{(\pm1,0\pm1)}&=\frac{1}{4}\rho(\Pi_x\pm u_x)(\Pi_z\pm u_z).\\
\end{split}
\end{align}
It is easy to verify by a direct computation that the moments of the populations (\ref{eq:D3Q19}) satisfy the relations given by the second row of Table \ref{table:PP}. Note that, when setting $\Pi_z=0$ in the nine populations, $f^*_{(0,0,0)}$, $f^*_{(\pm1,0,0)}$, $f^*_{(0,\pm1,0)}$, and $f^*_{(\pm1,\pm1,0)}$ (\ref{eq:D3Q19}) we again obtain the UniQuE on the two-dimensional D2Q9 lattice (\ref{eq:UniQuED2Q9}).
Maxwellization (\ref{eq:MPP}) of (\ref{eq:D3Q19}) is achieved upon substitution of (\ref{eq:MaxwellizatorD1Q3}):
\begin{align}
\begin{split}\label{eq:MD3Q19}
{f}^{\rm M}_{(0,0,0)}&=\frac{1}{3}\rho\left[1-(u_{x}^2+u_{y}^2+u_{z}^2)+3(u_x^2u_y^2+u_y^2u_z^2+u_x^2u_z^2)\right],\\
{f}^{\rm M}_{(\pm1,0,0)}&=\frac{1}{18}\rho
\left[1-3(u_y^2+u_z^2)\right]\left(1\pm 3u_x+3u_{x}^2\right),\\
{f}^{\rm M}_{(0,\pm1,0)}&=\frac{1}{18}\rho
\left[1-3(u_x^2+u_z^2)\right]\left(1\pm 3u_y+3u_{y}^2\right),\\
{f}^{\rm M}_{(0,0,\pm1)}&=\frac{1}{18}\rho\left[1-3(u_x^2+u_y^2)\right]\left(1\pm 3u_z+3u_{z}^2\right),\\
{f}^{\rm M}_{(\pm1,\pm1,0)}&=\frac{1}{36}\rho \left(1\pm 3u_x+3u_{x}^2\right)\left(1\pm 3u_y+3u_{y}^2\right),\\
{f}^{\rm M}_{(0,\pm1,\pm1)}&=\frac{1}{36}\rho \left(1\pm 3u_y+3u_{y}^2\right)\left(1\pm 3u_z+3u_{z}^2\right),\\
{f}^{\rm M}_{(\pm1,0\pm1)}&=\frac{1}{36}\rho \left(1\pm 3u_x+3u_{x}^2\right)\left(1\pm 3u_z+3u_{z}^2\right).\\
\end{split}
\end{align}
Positivity domain of (\ref{eq:MD3Q19}) is the intersection of three cylinders, $C_x=\{\bm{u}: u_y^2+u_z^2<\frac{1}{3}\}$,  $C_y=\{\bm{u}: u_x^2+u_z^2<\frac{1}{3}\}$ and $C_z=\{\bm{u}: u_x^2+u_y^2<\frac{1}{3}\}$:
$\Omega_{\rm D3Q19}^{\rm M}=C_x\bigcap C_y\bigcap C_z$. Since $\Omega_{\rm D3Q19}^{\rm M}$ is included in a box with the edge $2/\sqrt{3}$, we have the following inclusion relations between the positivity domains:
\begin{equation}
\Omega_{\rm D3Q19}^{\rm M}\subset\Omega_{\rm D3Q27}^{\rm M}\subset\Omega_{\rm D3Q27}^{\rm E}.
\end{equation}
Although the positivity domain shrinks when proceeding from the Maxwell to the sub-Maxwell lattice, all the three equilibria are well consistent with the low Mach number restriction to these models, $|u_{\alpha}|\ll 1/\sqrt{3}$.
Functions (\ref{eq:D3Q19}) and (\ref{eq:MD3Q19}) are used in \cite{Asinari09a} for the construction of a three-dimensional two relaxation time LB model.

\blue{\section{Quasi-equilibrium D3Q13 model}\label{appendix:D3Q13}}

The D3Q13 is the sub-Maxwell lattice of the D3Q27 with the smallest number of velocities capable of retaining the pressure tensor.
The peculiarity of the D3Q13 as compared to the other lattices (the Maxwell D3Q27 and the sub-Maxwell D3Q15 and D3Q19 lattices) is in the third-order moment tensor $Q_{\alpha\beta\gamma}$. Indeed, the D3Q27, D3Q15 and D3Q19 lattices all recover the isotropic linear part of the equilibrium function $Q^{\rm M}_{\alpha\beta\gamma}$ in the form
\begin{equation}\label{eq:ThirdMomentCorrect}
Q^{\rm M}_{\alpha\beta\gamma}=\frac{1}{3}(u_\alpha\delta_{\beta\gamma}+u_\beta\delta_{\alpha\gamma}+u_\gamma\delta_{\alpha\beta})+O(u^3),
\end{equation}
which corresponds to the linear in $u$ piece of the correct Maxwell moment relation at the reference temperature $T_0=1/3$. Terms of order $O(u^3)$ are different for each of the D3Q27, D3Q15 or D3Q19 lattices but their effect is negligible at low Mach numbers.
On the contrary, the corresponding expression for D3Q13 is not isotropic even at the linear order:
\begin{align}
\begin{split}\label{eq:peculiarityD3Q13}
Q^{\rm M}_{\alpha\alpha\alpha}&=u_\alpha,\\
Q^{\rm M}_{\alpha\beta\beta}&=\frac{1}{2}u_\alpha+O(u^3),\ \alpha\ne\beta,\\
Q^{\rm M}_{xyz}&=O(u^3).
\end{split}
\end{align}
Note that the factor $1/2$ instead of $1/3$ in the off-diagonal terms $Q^{\rm M}_{\alpha\beta\beta}$ (\ref{eq:peculiarityD3Q13}) is inconsistent with the correct Maxwell relation (\ref{eq:ThirdMomentCorrect}) (in other words, the diagonal terms $Q^{\rm M}_{\alpha\alpha\alpha}$ in (\ref{eq:peculiarityD3Q13}) correspond to the correct reference temperature $T_0=1/3$ whereas the off-diagonal terms $Q^{\rm M}_{\alpha\beta\beta}$ correspond to a different "temperature" $1/2$). Thus, the D3Q13 lattice is less isotropic than any of the other sub-Maxwell lattices (D3Q15 or D3Q19) of the Maxwell D3Q27 lattice. This peculiarity precludes developing the standard LBGK model on the D3Q13 lattice, as was first noticed in \cite{dHumieres01} upon a different consideration.

Utilizing the concept of UniQuE, we shall now derive a simple BGK-like model with two relaxation times which recovers the incompressible Navier-Stokes equations on the D3Q13 lattice.
For that, we use a generic pattern of quasi-equilibrium kinetic equations with a two-step relaxation mechanism \cite{Gorban94,Levermore96,Ansumali07,Asinari09,Asinari09a},
\begin{equation}
\label{eq:two-stepQE}
\partial_t f+\bm{v}\cdot\nabla f=-\frac{1}{\tau_1}(f-f^*)-\frac{1}{\tau_2}(f^*-f^{\rm M}),
\end{equation}
where the first term in the right hand side describes a relaxation to the UniQuE state $f^*$ (with a rate $\tau_1$), and the second term represents a relaxation from the UniQuE to the equilibrium (with a rate $\tau_2$). A rationale behind using a two-step quasi-equilibrium model (\ref{eq:two-stepQE}) in the present context is the following: The two steps of relaxation "adjust" separately the off-diagonal and the diagonal components of the nonequilibrium pressure tensor (see below) and will be tailored in such a way as to recover isotropy in the low Mach number limit (see also Refs. \cite{Ansumali07,Asinari09} for the application of this type of models in various other context).
For the present case, we choose the UniQuE of the projection pruning (see Tab.\ \ref{table:PP}), and the equilibrium as the Maxwellization thereof.
Specifically,
%
%
the UniQuE (\ref{eq:UniQuEPP}) is found by an inversion of the moment relations given in the D3Q13 row of Tab.\ \ref{table:PP}. It proves convenient to
restore a notation $\Pi_{\alpha\alpha}=\Pi_{\alpha}$ for the diagonal components of the pressure tensor at unit density:
\begin{align}
\begin{split}\label{eq:UniQuED3Q13}
{f}^*_{(0,0,0)}&=\rho\left(1-\frac{1}{2}(\Pi_{xx}+\Pi_{yy}+\Pi_{zz})\right),\\
f^*_{(\sigma,\lambda,0)}&=\frac{1}{8}\rho\left[(\Pi_{xx}+\sigma u_{x})(1+\lambda u_y)+(\Pi_{yy}+\lambda u_y)(1+\sigma u_x)-\Pi_{zz}(1+\sigma u_x+\lambda u_y)\right],\\
f^*_{(\sigma,0,\lambda)}&=\frac{1}{8}\rho\left[(\Pi_{xx}+\sigma u_x)(1+\lambda u_z)+(\Pi_{zz}+\lambda u_z)(1+\sigma u_x)-\Pi_{yy}(1+\sigma u_x+\lambda u_z)\right],\\
f^*_{(0,\sigma,\lambda)}&=\frac{1}{8}\rho\left[(\Pi_{yy}+\sigma u_y)(1+\lambda u_z)+(\Pi_{zz}+\lambda u_z)(1+\sigma u_y)-\Pi_{xx}(1+\sigma u_y+\lambda u_z)\right],\\
\end{split}
\end{align}
where $\sigma\in\{-1,1\}$ and $\delta\in\{-1,1\}$, and the Maxwellization (\ref{eq:MPP}) is achieved upon substituting $\Pi_{\alpha\alpha}^{\rm M}=1/3+u_{\alpha}^2$
(\ref{eq:MaxwellizatorD1Q3}) into the above expression (\ref{eq:UniQuED3Q13}).

It can be shown that, if the relaxation times $\tau_1$ and $\tau_2$ are chosen as
\begin{equation}\label{eq:taus}
\tau_1=\tau,\ \tau_2=2\tau,\ \tau>0,
\end{equation}
then, under the diffusive scaling at low Mach number ($\partial_t\to\epsilon^2\partial_t$, $\partial_x\to\epsilon\partial_x$, $\bm{u}=\epsilon\bm{u}^{(1)}$, $\rho=1+\epsilon^2 \rho^{(2)}$, where $\epsilon$ is the Mach number, see, e.\ g.\ \cite{Sone02,Lebowitz89incompressible,Bardos91,Junk05,AsinariOhwada09}), kinetic equation (\ref{eq:two-stepQE}) reduces to the incompressible Navier-Stokes equation,
\begin{eqnarray}
&&\nabla\cdot\bm{u}^{(1)}=0, \label{eq:solenoidal}\\
&&\partial_t \bm{u}^{(1)}+\bm{u}^{(1)}\cdot\nabla\bm{u}^{(1)}+\nabla p-\nu\Delta \bm{u}^{(1)}=0,\label{eq:NS}
\end{eqnarray}
where $\Delta=\partial_x^2+\partial_y^2+\partial_z^2$ is Laplace operator, $p=\rho^{(2)}/3$ is the hydrodynamic pressure defined by the solenoidal (incompressibility) condition (\ref{eq:solenoidal}), and $\nu$ is the kinematic viscosity given by the formula,
\begin{equation}
\nu=\tau/2.
\end{equation}
Note that the kinetic model (\ref{eq:two-stepQE}) is realizable under the condition (\ref{eq:taus}): relaxation towards the quasi-equilibrium is faster than the relaxation from the quasi-equilibrium to the equilibrium ($\tau_1<\tau_2$).

The simplest way to prove this statement is to consider a closed moment system equivalent to the kinetic model (\ref{eq:two-stepQE}) for the moments
\begin{equation}
\rho{\cal{M}}=\{\rho, \rho u_x,\rho u_y,\rho u_z, \rho \Pi_{xx}, \rho \Pi_{yy}, \rho \Pi_{zz}, \rho \Pi_{xy}, \rho \Pi_{yz}, \rho \Pi_{xz},
\rho T_{x}, \rho T_{y}, \rho T_{z}\},
\end{equation}
where the three independent third-order moments $T_{\alpha}$ are defined as
\begin{align}
\begin{split}
\rho T_x=\left\langle v_{(i)}(v^2_{(j)}-v^2_{(k)})f_{(i,j,k)}\right\rangle,\\
\rho T_y=\left\langle v_{(j)}(v^2_{(i)}-v^2_{(k)})f_{(i,j,k)}\right\rangle,\\
\rho T_z=\left\langle v_{(k)}(v^2_{(i)}-v^2_{(j)})f_{(i,j,k)}\right\rangle.
\end{split}
\end{align}
The moment system equivalent to the kinetic equation (\ref{eq:two-stepQE}) reads
\begin{align}
\begin{split}
\label{eq:D3Q13momsys:continuity}
&\epsilon^2 \partial_t \rho +\epsilon\partial_x (\rho u_x) +\epsilon\partial_y (\rho u_y)+\epsilon\partial_z(\rho u_z)=0,\\
\end{split}
\end{align}
\begin{align}
\begin{split}
\label{eq:D3Q13momsys:momentum}
&\epsilon^2\partial_t (\rho u_x)+ \epsilon\partial_x(\rho\Pi_{xx})+\epsilon\partial_y(\rho\Pi_{xy})+\epsilon\partial_z(\rho \Pi_{xz})=0,\\
&\epsilon^2\partial_t (\rho u_y)+ \epsilon\partial_x(\rho\Pi_{xy})+\epsilon\partial_y(\rho\Pi_{yy})+\epsilon\partial_z(\rho \Pi_{yz})=0,\\
&\epsilon^2\partial_t (\rho u_z)+ \epsilon\partial_x(\rho\Pi_{xz})+\epsilon\partial_y(\rho\Pi_{yz})+\epsilon\partial_z(\rho \Pi_{zz})=0,\\
\end{split}
\end{align}
\begin{align}
\begin{split}
\label{eq:D3Q13momsys:diagonal}
&\epsilon^2\partial_t (\rho \Pi_{xx})+\epsilon\partial_x(\rho u_x)+\frac{1}{2}\epsilon\partial_y(\rho(u_y+T_y) )+\frac{1}{2}\epsilon\partial_z (\rho (u_z+T_z))
=-\frac{1}{\tau_2}\rho\left(\Pi_{xx}-\left(\frac{1}{3}+u_x^2\right)\right),\\
&\epsilon^2\partial_t (\rho \Pi_{yy})+\epsilon\partial_y(\rho u_y)+\frac{1}{2}\epsilon\partial_x(\rho(u_x+T_x) )+\frac{1}{2}\partial_z (\rho (u_y+T_y))=
-\frac{1}{\tau_2}\rho\left(\Pi_{yy}-\left(\frac{1}{3}+u_y^2\right)\right),\\
&\epsilon^2\partial_t (\rho \Pi_{zz})+\epsilon\partial_z(\rho u_z)+\frac{1}{2}\epsilon\partial_y(\rho(u_y+T_y) )+\frac{1}{2}\epsilon\partial_z (\rho (u_z+T_z))=
-\frac{1}{\tau_2}\rho\left(\Pi_{zz}-\left(\frac{1}{3}+u_z^2\right)\right),\\
\end{split}
\end{align}
\begin{align}
\begin{split}
\label{eq:D3Q13momsys:offdiagonal}
&\epsilon^2\partial_t (\rho \Pi_{xy})+\frac{1}{2}\epsilon\partial_x (\rho(u_y+T_y) )+\frac{1}{2}\epsilon\partial_y (\rho (u_x+T_x))=-\frac{1}{\tau_1}\rho(\Pi_{xy}- u_x u_y),\\
&\epsilon^2\partial_t (\rho \Pi_{yz})+\frac{1}{2}\epsilon\partial_y(\rho(u_z+T_z) )+\frac{1}{2}\epsilon\partial_z (\rho (u_y+T_y))=-\frac{1}{\tau_1}\rho(\Pi_{yz}- u_yu_z),\\
&\epsilon^2\partial_t (\rho \Pi_{xz})+\frac{1}{2}\epsilon\partial_x (\rho(u_z+T_z) )+\frac{1}{2}\epsilon\partial_z (\rho (u_x+T_x))=-\frac{1}{\tau_1}\rho(\Pi_{xz}- u_x u_z),\\
\end{split}
\end{align}
\begin{align}
\begin{split}
\label{eq:D3Q13momsys:T}
\epsilon^2\partial_t(\rho T_x)+\epsilon\partial_x(\rho(\Pi_{yy}-\Pi_{zz}))+\epsilon\partial_y(\rho\Pi_{yz})-\epsilon\partial_z(\rho\Pi_{xz})=&
-\frac{1}{\tau_1}\rho(T_x-u_x(\Pi_{yy}-\Pi_{zz}))\\
&-\frac{1}{\tau_2}\rho u_x(\Pi_{yy}-\Pi_{zz}-u_y^2+u_z^2),\\
\epsilon^2\partial_t(\rho T_y)+\epsilon\partial_y(\rho(\Pi_{xx}-\Pi_{zz}))+\epsilon\partial_x(\rho\Pi_{xy})-\epsilon\partial_z(\rho\Pi_{yz})&=
-\frac{1}{\tau_1}\rho(T_y-u_y(\Pi_{xx}-\Pi_{zz}))\\
&-\frac{1}{\tau_2}\rho u_y(\Pi_{xx}-\Pi_{zz}-u_x^2+u_z^2),\\
\epsilon^2\partial_t(\rho T_z)+\epsilon\partial_z(\rho(\Pi_{xx}-\Pi_{yy}))+\epsilon\partial_x(\rho\Pi_{xz})-\epsilon\partial_y(\rho\Pi_{yz})=&
-\frac{1}{\tau_1}\rho(T_z-u_z(\Pi_{xx}-\Pi_{yy}))\\
&-\frac{1}{\tau_2}\rho u_z(\Pi_{xx}-\Pi_{yy}-u_x^2+u_y^2),\\
\end{split}
\end{align}
where we have explicitly introduced the diffusion scaling. Substituting $\rho=1+\epsilon^2 \rho^{(2)}$ and $\bm{u}=\epsilon\bm{u}^{(1)}$
into the continuity equation (\ref{eq:D3Q13momsys:continuity}), we find at the first non-trivial order:
\begin{equation}\label{eq:solenoidalcondition}
\epsilon^2\partial_\alpha{u}^{(1)}_\alpha=0,
\end{equation}
where summation convention is applied.
Next, from the relaxation equations for the components of the pressure tensor, Eqs.\ (\ref{eq:D3Q13momsys:diagonal}) and (\ref{eq:D3Q13momsys:offdiagonal}), it follows that
\begin{equation}
\rho\Pi_{\alpha\beta}=\frac{1}{3}\delta_{\alpha\beta}+\epsilon^2\left(\frac{1}{3}\delta_{\alpha\beta}\rho^{(2)}+u^{(1)}_\alpha u^{(1)}_\beta\right)+\epsilon^2 \Pi_{\alpha\beta}^{{\rm neq}(2)}+O(\epsilon^3),
\end{equation}
where the term $\Pi_{\alpha\beta}^{{\rm neq}(2)}$ is the non-equilibrium (viscous) part of the pressure tensor which is not yet defined.
Substituting the latter expression into the momentum equation (\ref{eq:D3Q13momsys:momentum}) yields
\begin{equation}\label{eq:pre-momentum}
\epsilon^3\left(\partial_t u^{(1)}_{\alpha}+\partial_{\alpha}(\rho^{(2)}/3)+u^{(1)}_{\beta}\partial_{\beta}u^{(1)}_\alpha +\partial_{\beta}\Pi_{\alpha\beta}^{{\rm neq}(2)}\right)=0,
\end{equation}
where we have made use of the solenoidal condition (\ref{eq:solenoidalcondition}). What remains is to derive the nonequilibrium part $\Pi_{\alpha\beta}^{{\rm neq}(2)}$. For that, let us consider again the moment equations for the components of the pressure tensor.
These give:
\begin{equation}\label{offdiagonalP}
\epsilon^2\frac{1}{2}\left(\partial_x u^{(1)}_y + \partial_y u^{(1)}_x\right)=-\frac{1}{\tau_1}\epsilon^2 \Pi_{xy}^{{\rm neq}(2)},
\end{equation}
for the off-diagonal component $\Pi_{xy}$ and similarly for the rest of the off-diagonal components, Eq.\ (\ref{eq:D3Q13momsys:offdiagonal}), and
\begin{equation}\label{eq:diagonalP}
\epsilon^2\left(\partial_x u^{(1)}_x+\frac{1}{2}\left(\partial_y u^{(1)}_y + \partial_z u^{(1)}_z\right)\right)=-\frac{1}{\tau_2}\epsilon^2 \Pi_{xx}^{{\rm neq}(2)},
\end{equation}
for the diagonal component $\Pi_{xx}$ and similarly to other diagonal components, Eq.\ (\ref{eq:D3Q13momsys:diagonal}).
Note that, when deriving the above results, we have used the property of the third-order moments, $T_{\alpha}=O(\epsilon^3)$, which follows from
the right hand side of the moment equations for $T_\alpha$ (Eq.\ (\ref{eq:D3Q13momsys:T})). Using once again the solenoidal condition (\ref{eq:solenoidalcondition}), Eq.\ (\ref{eq:diagonalP}) can be rewritten:
\begin{equation}
\frac{1}{4}\epsilon^2\left(\partial_x u^{(1)}_x+\partial_x u^{(1)}_x\right)=-\frac{1}{\tau_2}\epsilon^2 \Pi_{xx}^{{\rm neq}(2)}.
\end{equation}
Thus, by choosing the relaxation times as $\tau_1=\tau$, $\tau_2=2\tau$, the nonequilibrium pressure tensor becomes isotropic:
\begin{equation}\label{eq:Pisotropic}
\Pi_{\alpha\beta}^{{\rm neq}(2)}=-\frac{\tau}{2}\left(\partial_{\alpha}u^{(1)}_{\beta}+\partial_{\beta}u^{(1)}_{\alpha}\right).
\end{equation}
Substituting (\ref{eq:Pisotropic}) into the momentum equation (\ref{eq:pre-momentum}) concludes the derivation of the incompressible Navier-Stokes equations (\ref{eq:NS}) from the quasi-equilibrium kinetic model (\ref{eq:two-stepQE}).

Finally, it is straightforward to derive a lattice Boltzmann scheme for the kinetic equation (\ref{eq:two-stepQE}) following a general method of Refs.\ \cite{Ansumali07,Asinari09,Asinari09a}:
Kinetic equation (\ref{eq:two-stepQE})
is integrated in time from $t$ to $t+\delta t$ along characteristics,
and the time integral of the right hand side, $J=-\frac{1}{\tau_1}(f-f^*)-\frac{1}{\tau_2}(f^*-f^{\rm M})$, is
evaluated by trapezoidal rule to get
\begin{equation}
\label{cn}
f(\bm{x}+\bm{v}\delta t,t+\delta t)-f(\bm{x},t)=
\frac{\delta t}{2}\,J(f(\bm{x}+\bm{v}\delta t,t+\delta t))
+\frac{\delta t}{2}\,J(f(\bm{x},t)).
\end{equation}
In order to avoid implicit computations in the latter expression, let us apply the following variable transform \cite{he98,Ansumali07}:
\begin{equation}
f\to g=f-\frac{\delta t}{2}J(f),
\end{equation}
to Eq. (\ref{cn}), which yields after taking into account (\ref{eq:taus}):
\begin{equation}
\label{tvlb}
g(\bm{x}+\bm{v}\delta t,t+\delta t)=(1-\omega)g(\bm{x},t)+\frac{\omega}{2}\left[f^{\rm M}(\rho,\bm{u})
+ f^*(\rho,\bm{u},\Pi_{\alpha\alpha}')\right],
\end{equation}
where
\begin{align}
\begin{split}
&\omega=\frac{2\delta t}{2\tau+\delta t},\\
&\rho=\rho(g),\\
&\bm{u}=\bm{u}(g),\\
&\Pi'_{\alpha\alpha}=\frac{1}{4\tau+\delta t}\left[4\tau\Pi_{\alpha\alpha}(g)+\delta t \Pi^{\rm M}_{\alpha\alpha}(g)\right].
\end{split}
\end{align}
The scheme (\ref{tvlb}) becomes the LB scheme if the time step $\delta t$ is matched with the lattice.
A different MRT LB equation for the D3Q13 lattice was suggested in \cite{dHumieres01}.
\end{appendix}
\end{document}